\documentclass[]{bytedance_seed}

\usepackage[toc,page,header]{appendix}
\usepackage[utf8]{inputenc} 
\usepackage[T1]{fontenc}    
\usepackage{hyperref}       
\usepackage{url}            
\usepackage{booktabs}       
\usepackage{amsfonts}       
\usepackage{nicefrac}       
\usepackage{microtype}      
\usepackage{threeparttable}
\usepackage{tabularx}
\usepackage{listings}
\usepackage{xcolor}
\usepackage{graphicx}
\usepackage{caption}
\usepackage{framed}
\usepackage{amssymb}
\usepackage{mathrsfs}
\usepackage{mathtools}
\usepackage{array}
\usepackage{amsthm}
\usepackage{verbatim} 
\usepackage{enumerate}
\usepackage{bbm}
\usepackage{commath}
\usepackage{wrapfig}
\usepackage{amsbsy}
\usepackage{float}
\usepackage{amsmath}
\usepackage{tabularx}
\usepackage{listings}
\usepackage{xcolor}
\usepackage{colortbl}
\usepackage{xltabular}
\usepackage{longtable}
\usepackage{booktabs}
\usepackage[shortlabels]{enumitem}
\usepackage{multirow}
\usepackage{algorithm}
\usepackage[algo2e]{algorithm2e} 
\usepackage{bbding}
\usepackage{pifont}
\usepackage{minted}
\usepackage{algorithmic}

\usepackage{amsmath}
\usepackage{amssymb}
\usepackage{mathtools}
\usepackage{amsthm}
\usepackage[table]{xcolor} 
\theoremstyle{plain}

\theoremstyle{definition}

\theoremstyle{remark}

\usepackage[textsize=tiny]{todonotes}

\newcolumntype{M}{>{$}c<{$}}

\newcommand{\model}{\textsc{IFCodeEvolve}\xspace}
\newcommand{\bench}{\textsc{IFCodeBench}\xspace}

\newcommand{\gtext}[1]{\textcolor{gray}{#1}}

\newsavebox{\codeboxPositive}
\newsavebox{\codeboxNegative}

\definecolor{mygray}{gray}{.85}
\definecolor{myyellow}{RGB}{204,102,0}
\definecolor{myred}{RGB}{204,0,102}
\definecolor{mypurple}{RGB}{102,0,204}
\definecolor{maroon}{cmyk}{0,0.87,0.68,0.32}
\definecolor{myblue}{RGB}{227,227,240}

\definecolor{PromptGray}{RGB}{120,120,120}
\definecolor{PromptBlue}{RGB}{48,98,166}
\definecolor{PromptBrown}{RGB}{139,94,60}
\definecolor{PromptTeal}{RGB}{46,125,122}

\tcbset{
  promptbox/.style={
    enhanced,
    colback=white,
    breakable,
    colframe=black!25,
    boxrule=2.0pt,
    arc=2pt,
    outer arc=2pt,
    left=6pt,right=6pt,top=6pt,bottom=6pt,
    fonttitle=\bfseries\small,
    coltitle=white,
    attach boxed title to top left={xshift=6pt,yshift=-1.5mm},
    boxed title style={
      boxrule=0pt,
      arc=2pt,
      outer arc=2pt,
      left=5pt,right=5pt,top=2pt,bottom=2pt
    },
  }
}

\newtcolorbox{PromptGrayBox}[1]{promptbox,
  title=#1,
  colframe=PromptGray,
  boxed title style={colback=PromptGray}
}

\newtcolorbox{PromptBlueBox}[1]{promptbox,
  title=#1,
  colframe=PromptBlue,
  boxed title style={colback=PromptBlue}
}

\newtcolorbox{PromptBrownBox}[1]{promptbox,
  title=#1,
  colframe=PromptBrown,
  boxed title style={colback=PromptBrown}
}

\newtcolorbox{PromptTealBox}[1]{promptbox,
  title=#1,
  colframe=PromptTeal,
  boxed title style={colback=PromptTeal, colframe=PromptTeal}
}

\definecolor{codebg}{gray}{0.95}

\lstdefinestyle{python}{
    language=Python,
    basicstyle=\ttfamily\small,
    keywordstyle=\color{blue},
    stringstyle=\color{teal},
    commentstyle=\color{gray},
    showstringspaces=false,
    breaklines=true,
    frame=none,
    backgroundcolor=\color{lightgray!10},
    xleftmargin=1.5em,
    columns=flexible
}

\hypersetup{
    colorlinks,
    linkcolor={red!50!black},
    citecolor={blue!50!black},
    urlcolor={blue!80!black}
}

\usepackage{minitoc}

\title{Steerable Instruction Following Coding Data Synthesis with Actor-Parametric Schema Co-Evolution}

\author[1,*]{Tinglin Huang}
\author[2,*,\dagger]{Bo Chen}
\author[2]{Xiao Zhang}
\author[2]{Kai Shen}
\author[1]{Rex Ying}

\affiliation[1]{Yale University}
\affiliation[2]{ByteDance Seed}

\contribution[*]{Equal Contribution}
\contribution[\dagger]{Corresponding authors}

\abstract{
Interpreting and following human instructions is a critical capability of large language models (LLMs) in automatic programming. 
However, synthesizing large-scale instruction-paired coding data remains largely unexplored and is particularly challenging when ensuring logical compatibility among multiple constraints.
In this study, we propose \model, an actor-schema co-evolution framework for instruction following coding data generation.
By representing instructions as parametric function schema, we construct a library that covers the vast instruction space via dynamic constraint instantiation.
Building upon this, Monte Carlo Tree Search (MCTS) sampler is applied to efficiently navigate this space, utilizing actor model feedback as a dynamic termination signal. 
Furthermore, to progressively explore challenging problems, we introduce a co-evolving paradigm that iteratively advances both the actor model and the schema library, via schema composition and mutation, based on sampler statistics. 
Empirical results demonstrate that \model significantly boosts base model performance, with our 32B model achieving parity with proprietary SOTA models. Additionally, we contribute \bench, a comprehensive human-verified benchmark equipped with solutions and robust AST-based verification.
}

\date{\today}
\correspondence{\email{chenbo.1015@bytedance.com}}

\begin{document}
\maketitle


\section{Introduction}

\begin{wrapfigure}{r}{0.45\textwidth}
    \centering
    \vspace{-15pt}
    \includegraphics[width=0.44\textwidth]{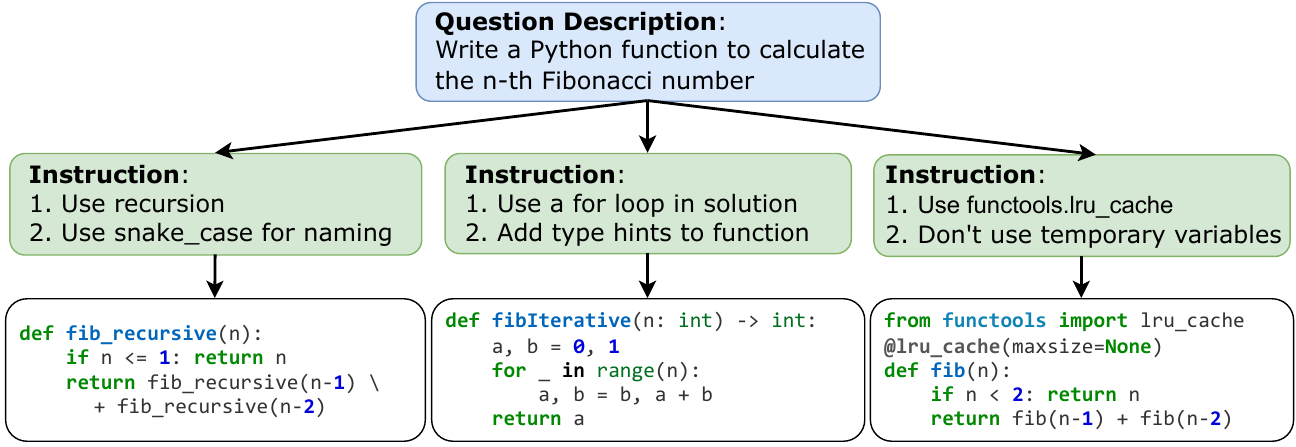}
    \caption{Instruction-driven code generation.}
    \label{fig:teaser}
    \vspace{-10pt}
\end{wrapfigure}

Large language models~(LLMs) have recently achieved remarkable success in automatic programming, spanning domains from competitive programming~\cite{li2022competition,achiam2023gpt,seed2025seed} to software engineering~\cite{jimenez2023swe,zan2025multi,guo2024deepseek}. 
Beyond fundamental coding proficiency, an important factor leading to these advancements is the instruction following~(IF) capability~\cite{ouyang2022training}. IF serves as a bridge between human intent expressed in natural language and executable logic,
as illustrated in Figure~\ref{fig:teaser}.

Improving the IF capability of LLMs necessitates large-scale coding data paired with high-quality instructions, which currently depends heavily on manual curation. 
While recent approaches have explored automatic data synthesis using static brute-force sampling~\cite{xu2024wizardlm,dong2024self} or adversarial self-evolving frameworks~\cite{zhao2025absolute,huang2025r}, they remain limited by two challenges:
\textbf{(1) Validity and verification bottleneck}: 
Directly prompting LLMs to generate multi-constraint coding problems struggles to guarantee logical consistency and solvability~\cite{yan2025codeif,yang2025ifevalcode}. 
Furthermore, generating accurate verification logic (e.g., unit tests) is itself error-prone.
\textbf{(2) Unstructured hard-sample exploration}: 
Static frameworks rely primarily on the generator’s intrinsic priors, resulting in repetitive data patterns and limited coverage of out-of-distribution samples.
In contrast, without carefully engineered and highly intricate reward designs, the generator of adversarial strategies can easily hack the reward by producing unsolvable or ill-defined problems.


\begin{figure}[t]
    \centering
    \includegraphics[width=1.0\textwidth]{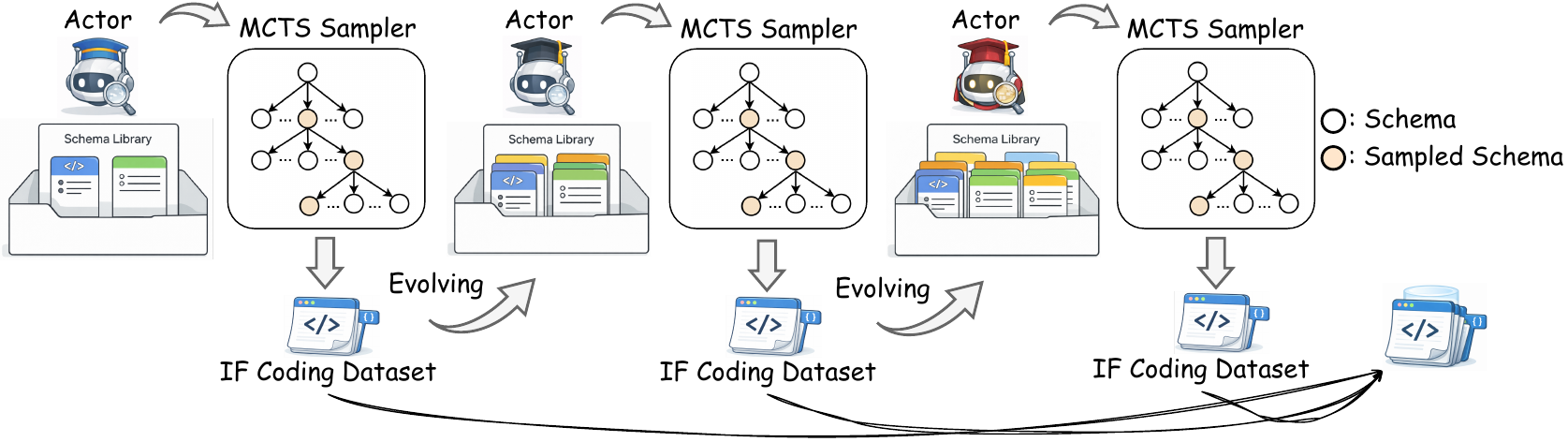}
    \caption{
        Illustration of actor-schema co-evolution paradigm for IF coding data generation.
    }
    \label{fig:co-evolving}
    \vspace{-0.2cm}
\end{figure}

Motivated by this, we propose \model, a framework that optimizes logical validity and problem complexity by unifying progressive proof-by-construction and adversarial actor-schema co-evolution.
We apply parametric instructions, which represents a vast instruction space within a compact library of schema. 
This formulation transforms data synthesis into a steerable and sequential sampling process. 
Specifically, we frame synthesis as MCTS trajectory sampling integrated with proof-by-construction validation strategy: 
at each step, the sampler navigates the combinatorial schema space to incrementally impose constraints, which are immediately instantiated and verified via code adaptation to ensure strict solvability.

Furthermore, this structured representation facilitates the exploration of high-complexity problems through actor-schema co-evolution: in parallel with the actor iteratively improves via post-training, the schema library autonomously evolves by merging synergistic primitives and mutating under-performing constraints. 
Unlike methods that improve data distribution by explicitly training generator, 
we enhance data synthesis in a steerable manner, explicitly optimizing structural primitives to continuously challenge the evolving model.
The overall paradigm is illustrated in Figure~\ref{fig:co-evolving}, where the sampler constructs IF coding data with iteratively improved actor and instruction schema.

We evaluate \model on two public benchmarks across five LLMs with varying parameter sizes (1.3B to 32B). 
Empirical results demonstrate that our framework drives consistent performance gains to the actor models across evolving procedures. 
Notably, the larger models trained on our synthesized data achieve a substantial improvement, achieving performance comparable to proprietary SOTA reasoning models.
To complement these general baselines with fine-grained assessment, we introduce \bench, a curated benchmark equipped with executable tests and AST-based verification protocols. 
Finally, we posit that this paradigm, which extends the scope of evolution from the target model to the core components of the synthesis framework itself, offers a generalizable methodology for constructing self-adapting curricula in other reasoning-intensive domains, such as mathematical problem solving~\cite{shao2024deepseekmath,xin2025bfs} and scientific discovery~\cite{jumper2021highly,du2025accelerating}.

\section{Related Work}

\paragraph{Instruction Tuning}
Instruction-following~(IF) is a foundational capability of LLMs, requiring the model to accurately interpret complex instructions and synthesize valid responses within an open-ended search space~\cite{lou2023comprehensive}.
To address the scarcity of high-quality instruction data, prior studies have resorted to manual annotation~\cite{ouyang2022training,chung2024scaling} or leveraged LLMs for large-scale data generation based on seed datasets~\cite{wang2023self,xu2024wizardlm,dong2024self}, typically necessitating dedicated data filtering procedures. 
Concurrently, numerous benchmarks have been proposed to evaluate these capabilities across diverse domains, including single-turn QA~\cite{zhou2023instruction,pyatkin2025generalizing,jiang2024followbench}, multi-turn conversation~\cite{deshpande2025multichallenge}, and code generation~\cite{yan2025codeif,yang2025ifevalcode}.
Different from a static pipeline, our work introduces a co-evolutionary framework for IF coding data generation, which dynamically adapts its core synthesis modules to iteratively refine the data distribution.

\paragraph{Self-Evolving Agents}
As LLMs demonstrate increasingly reasoning capabilities, a promising line of research investigates their potential to self-optimize agentic systems~\cite{gao2025survey}.
Existing studies generally fall into four categories in terms of applications: \textbf{(1) prompt tuning}, where LLMs function as meta-optimizers to iteratively rewrite instructions and in-context exemplars, reducing reliance on manual configurations~\cite{yang2023large,yuksekgonul2025optimizing};
\textbf{(2) workflow construction}; which involves representing agentic topologies as computational graphs or code and optimizing interaction structures to solve complex problems~\cite{hu2024automated,zhang2024aflow}; 
\textbf{(3) solution refinement}, in which the system engages in recursive feedback loops to critique, debug, and iteratively improve its own outputs~\cite{madaan2023self,lin2025se,novikov2025alphaevolve};
and \textbf{(4) data augmentation/generation}, where the models distill knowledge into datasets by synthesizing high-quality reasoning trajectories for self-training~\cite{jumper2021highly,silver2016mastering,zelikman2024star,xin2024deepseek}, or employ adversarial strategies to generate challenging training instances~\cite{zhao2025absolute,huang2025r}.
Distinct from these prior approaches, \model implements a meta-level evolution of the data generation mechanism itself. By iteratively evolving the instruction schema and actor, we sustain an adaptive curriculum for coding data generation.

\section{Method}
In this section, we introduce \model, a co-evolving instruction-following coding data generation framework. We first formulate the problem and objective in Section~\ref{sec:problem_formulation}, and elaborate the core components in Section~\ref{sec:instruction_sampler}. The implementation of the proposed co-evolving data synthesis framework is presented in Section~\ref{sec:inner_loop} and Section~\ref{sec:outer_loop}.
The pesudocode can be found in Algo.\ref{algo:framework}.

\subsection{Problem Formulation}\label{sec:problem_formulation}

Given a set of coding problems, solutions, and functional unit tests $\mathcal{D}=\{(p_k,s_k,f_{\text{test}}^k(\cdot))\}^N_{k=1}$, where each coding problem $p_k$ includes a question description, function signature, and input/output example, the goal of our proposed framework is to augment each instance by synthesizing a set of instruction constraints:
\begin{align}
p'_k,s'_k,\boldsymbol{\iota}_{k}=f_{\text{IFCoEvolve}}(p_k,s_k)
\end{align}
where $\boldsymbol{\iota}_k=\{\iota_{k,j}\}^{M_k}_{j=1}$ is the set of instructions for $k$-th problem. 
We focus on verifiable instructions, i.e., the instructions can be automatically validated via checking functions, such as the variable name convention, data structure usage, or interface design. 
The instructions impose strict constraints without altering the core algorithmic semantics, enforcing the model to reason within a narrowed solution space.

\paragraph{Challenges}
Constructing high-quality IF coding datasets presents two primary challenges:
\textbf{(1) Constraint compatibility and solvability}: Unlike general chat instructions, coding constraints must be logically consistent with both the problem context and each other.
For instance, a restriction on variable naming conventions can directly conflict with a directive like ``must not use intermediate variables''.
Ensuring that the aggregated instructions remain soluble without creating logical conflict is non-trivial. 
\textbf{(2) Balancing difficulty with validity}: The generated instructions must be sufficiently challenging to push the model's reasoning boundary, targeting problems where the model currently fails. 
However, generating fundamentally unsolvable problems would easily satisfy such a criterion while hack the objective.
The core challenge lies in generating samples that are adversarially hard yet strictly valid.

\subsection{Instruction Representation and Sampling}\label{sec:instruction_sampler}

We introduce parametric instruction schema for steerable data synthesis, and an MTCS sampler used to explore instruction combinations from the vast schematic space.


\paragraph{Parametric Instruction}
Previous methods~\cite{dong2024self,xu2024wizardlm} prompt LLMs to brainstorm instruction candidates using static text representations. 
Such representations treat parametric variations of the same logic (e.g., variable length $>5$ vs. $<5$) as distinct entities, leading to a redundant search space.

\begin{figure*}[t]
    \centering
    \includegraphics[width=1.0\textwidth]{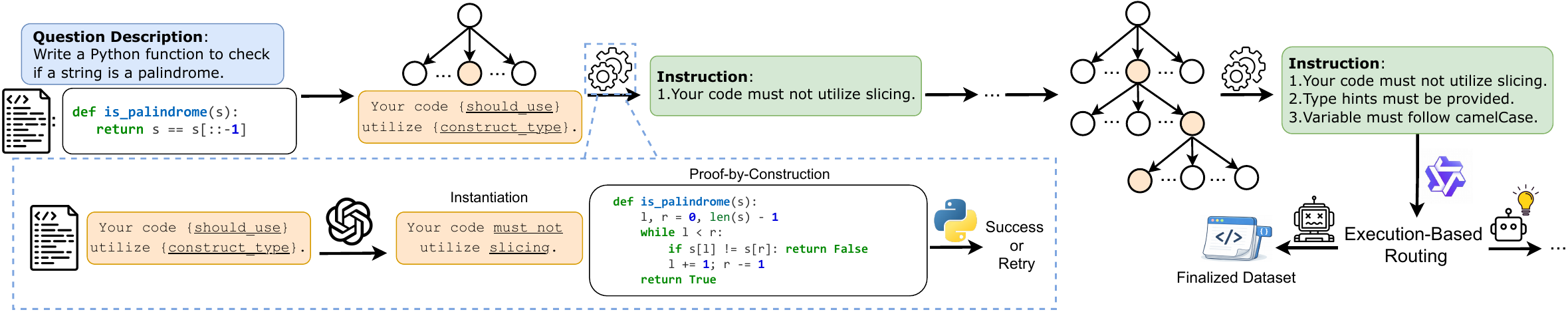}
    \caption{
        Illustration of multi-round IF coding data augmentation.
    }
    \label{fig:data_generation}
    \vspace{-0.2cm}
\end{figure*}

In light of this, we apply parametric instruction~\cite{yan2025codeif}, representing each instruction as an instantiable function schema, with configurable constraint attributes serving as parameters.
Formally, a concrete instruction $\iota$ is derived by instantiating a schema $\tau$ which consists of:
\begin{itemize}[leftmargin=*]
\item A textual skeleton with variable slots, e.g., ``The length of variable names \{\texttt{comparison}\} exceed \{\texttt{length}\} characters.''.
\item An argument space defining valid types and scope for the slots, e.g., \texttt{comparison} $\in [\text{``must''},\text{``must not''}]$ and \texttt{length} $\in$ int.
\item $f_{\text{IF-test}}(\cdot)$: An AST\footnote{\url{https://docs.python.org/3/library/ast.html}}-based verification function that maps a given solution to $\{0,1\}$, indicating whether it strictly adheres to the instantiated instruction.
\end{itemize}
The instantiation process is driven by an LLM conditioned on the problem context, as detailed in Section~\ref{sec:inner_loop}.
We construct an initial library $\mathcal{T}$ of 27 diverse schema
(see Appendix~\ref{app:dataset}), which serves as the foundation for the subsequent sampling procedure.

\paragraph{Monte Carlo Tree Search-based Sampler}
Building upon $\mathcal{T}$, we formulate the IF data synthesis process as a sequential tree search structure,
where each node expansion corresponds to instantiating a schema and a root-to-leaf trajectory constitutes the final aggregated instruction set.
To navigate this vast combinatorial space, we employ the sampler based on Monte Carlo Tree Search~(MCTS)~\cite{guo2025deepseek}, 
which manages the sampling by quantifying the effectiveness of schema, i.e., estimating the potential of a branch to yield a challenging instruction set.

Formally, let $\boldsymbol{\tau}_{<t}=\{\tau_1,\cdots,\tau_t\}$ denote the state at step $t$, representing the sequence of schema sampled so far. 
Each schema corresponds to an instantiated instruction, forming the set $\boldsymbol{\iota}_{<t}=\{\iota_1,\cdots,\iota_t\}$.
At each step, the MCTS selects a small batch of the schema from the remaining library $\mathcal{T}\setminus \boldsymbol{\tau}_{<t}$ based on the estimated value:
\begin{equation}\label{equ:sampling}
\{\tau\}^{C} \sim \underset{\tau \in \mathcal{T} \setminus \boldsymbol{\tau}_{<t}}{\operatorname{softmax}} \ Q(\boldsymbol{\tau}_{<t}, \tau) 
\end{equation}
where $C$ is the candidate size.
The value is calculated by aggregating the exploitation and exploration terms:
\begin{equation}
Q(\boldsymbol{\tau}_{<t}, \tau) = \frac{W(\boldsymbol{\tau}_{<t}, \tau) + \alpha \cdot P(\tau)}{N(\boldsymbol{\tau}_{<t}, \tau) + \alpha} + U(\boldsymbol{\tau}_{<t}, \tau)
\end{equation}
where $W(\boldsymbol{\tau}_{<t}, \tau)$ and $N(\boldsymbol{\tau}_{<t}, \tau)$ are accumulated success count and visit count of schema $\tau$ at state $\boldsymbol{\tau}_{<t}$, $P(\tau)$ represents the global prior success rate of $\tau$, and $\alpha$ is a smoothing factor (an integer $>1$) to mitigate high variance in low-visit nodes~\cite{zhai2017study}.
The exploration term $U(\boldsymbol{\tau}_{<t}, \tau)$ is calculated as:
\begin{equation}
U(\boldsymbol{\tau}_{<t}, \tau) = 
\begin{cases} 
\eta & \text{if } N(\boldsymbol{\tau}_{<t}, \tau) = 0, \\
\eta \sqrt{\frac{\ln N(\boldsymbol{\tau}_{<t})}{N(\boldsymbol{\tau}_{<t}, \tau)}} & \text{if } N(\boldsymbol{\tau}_{<t}, \tau) > 0.
\end{cases}
\end{equation}
where $N(\boldsymbol{\tau}_{<t})$ is the state visit count and $\eta$ is the exploration weight.
For a schema $\tau$ that has never been sampled at the current state, i.e., $N(\boldsymbol{\tau}_{<t}, \tau)=W(\boldsymbol{\tau}_{<t}, \tau)=0$, the value $Q(\boldsymbol{\tau}_{<t}, \tau)$ simplifies to $P(\tau)+\eta$, which prioritize schema with high global performance.

\paragraph{Discussion}

The utilization of parametric instructions enables us to cover a vast instruction space with a manageable set of schema. 
As such, it allows the application of MCTS, which is grounded in statistical priors to significantly enhance data synthesis efficiency.

\subsection{Multi-Round Augmentation}\label{sec:inner_loop}



As illustrated in Figure~\ref{fig:data_generation}, given a coding data and schema library, our framework iteratively adding one sampled instruction from MCTS to the problem,
validating solvability via a proof-by-construction instantiation and filtering instances through execution-based routing.


\paragraph{Proof-by-Construction Instantiation}
The dataset at step $t$ is denoted as $\mathcal{D}_t$, where each instance comprises the problem, the current solution, and the accumulated instructions, represented as $(p_t, s_t, \boldsymbol{\iota}_{<t})$.
For the subsequent step, MCTS samples a batch of candidate schema via Equ.~\ref{equ:sampling}. The transition to the next state is performed by an LLM:
\begin{equation}
p_{t+1},s_{t+1},\boldsymbol{\iota}_{<{t+1}}=\pi_\text{gen}(p_t,s_t,\boldsymbol{\iota}_{<t},\{\tau\}^C)
\end{equation}
where we consider the LLM serves as data generator $\pi_\text{gen}$.
Specifically, the LLM is prompted to rank the candidate schema based on the compatibility, subsequently instantiating the selected schema to maximize difficulty.

To guarantee validity, the LLM is required to synthesize a witness solution by modifying the current solution $s_t$ conditioned on the problem, the historical instruction set, and the newly added instruction. 
The instantiation is deemed successful only if the witness solution passes the cumulative set of IF verification functions as well as the functional unit tests, thereby demonstrating the solvability of the augmented problem.

Overall, instead of validating from scratch, we frame the augmentation as a progressive code adaptation task, injecting deterministic external feedback via execution to rigorously prove constraint feasibility. This step-wise verification enforces inductive solvability: by ensuring that every increment is valid, the final problem is guaranteed to be soluble.




\paragraph{Execution-Based Feedback Routing}
We implement a feedback-driven curriculum strategy by deploying an LLM as an actor model to dynamically probe the solvability of the generated data. 
The actor's execution pass rate serves as a real-time indicator of difficulty, which is calculated as:
\begin{align}
\left( \mathbb{E}_{s}\left[ f_{\text{test}}(s) \right] < \lambda \right) \land \left( \mathbb{E}_{s}\left[ \prod_{\iota} f_{\text{IF-test}}^\iota(s) \right] < \lambda \right)
\end{align}
where $\lambda$ denotes the difficulty threshold and solutions are drawn from the actor, i.e., $s \sim \pi_{\text{actor}}(\cdot|p_t, \boldsymbol{\iota}_{<t})$.
We estimate these expectations via Best-of-N sampling~\cite{wang2022self}.
This equation indicates that a problem is deemed difficult if both the average functional pass rate and the average IF all-pass rate fall below a specific threshold. 

The calculated indicator drives a conditional routing policy: 
after augmenting the dataset with new instructions at step $t$, 
instances that the actor fails to solve are early-stopped for finalization, whereas successfully solved instances are propagated to the next iteration for continued augmentation.
This indicator also functions as the win condition for MCTS. That is, if the current instruction can induce failure, it is counted as a win (i.e., $W(\boldsymbol{\tau}_{<t}, \tau) \leftarrow W(\boldsymbol{\tau}_{<t}, \tau) + 1$)).






\subsection{Co-Evolving Paradigm}\label{sec:outer_loop}

Our proposed augmentation paradigm establishes a robust pipeline for IF code data generation. 
However, adhering to a static schema library and a fixed actor limits problem diversity and difficulty at the actor's performance.
Motivated by this, we extend the static pipeline into a self-evolving framework, which facilitates the co-evolution of both the actor and the schema library, ensuring that the data generation process dynamically scales with the model's capabilities.

\paragraph{Actor Evolution}
In our framework, the actor serves as a dynamic filter, routing instances that it fails to solve into the finalized problem set. 
To ensure the generation of progressively challenging problems, we implement an iterative evolution paradigm: upon completing each augmentation round, the actor is post-trained on the newly synthesized data.
This optimized actor is used in the subsequent generation phase, raising the difficulty bar for future samples.

The evolving procedure of actor build a curriculum generation process: the problem patterns that the actor initially fails can potentially overcome post-training, prompting the paradigm to pivot towards generating strictly harder instances that remain unsolved by the improved actor.


\paragraph{Instruction Schema Evolution}
While the initial library $\mathcal{T}$ provides a foundation, a fixed schema set can let actor easily adapt to fixed patterns. 
In light of this, we propose to refine $\mathcal{T}$ via two strategies: composition and mutation.

As for the composition, we leverage MCTS trajectory statistics to identify synergistic schema compositions. Specifically, the global joint success rate for a transition $\tau_i \to \tau_j$ is aggregated over all visited states $\mathcal{S}$ ending in $\tau_i$:
\begin{equation}\label{equ:composition_1}
\bar{R}(\tau_i, \tau_j) = \frac{\sum_{\boldsymbol{\tau}_{<t} \in \mathcal{S}_{\tau_i}} W(\boldsymbol{\tau}_{<t}, \tau_j)}{\sum_{\boldsymbol{\tau}_{<t} \in \mathcal{S}_{\tau_i}} N(\boldsymbol{\tau}_{<t}, \tau_j)}
\end{equation}
where $\mathcal{S}_{\tau_i} = \{ \boldsymbol{\tau}_{<t} \mid \text{last}(\boldsymbol{\tau}_{<t}) = \tau_i \}$ denotes the subset of states where the most recently sampled schema is $\tau_i$.
A pair $(\tau_i, \tau_j)$ is selected for composition if its occurrence frequency demonstrates a synergistic difficulty gain compared to the individual priors $P(\cdot)$:
\begin{equation}\label{equ:composition_2}
\bar{R}(\tau_i, \tau_j) > \max \left( P(\tau_i), P(\tau_j) \right) + \delta
\end{equation}
where $\delta$ denotes the minimum margin of difficulty improvement. Selected pairs are merged to form a new composite schema $\tau_{i \oplus j}$, which unifies the logic of both constituents into a single parametric function.

As for the mutation, it targets the under-performing schema based on their global prior success rate $P(\tau)$, i.e., $\mathcal{T}_{\text{weak}} \subset \mathcal{T}$ with the lowest $P(\tau)$ values. 
For each $\tau \in \mathcal{T}_{\text{weak}}$, we retrieve a set of question instances including the instruction instantiated from $\tau$, where the actor successfully passes this question in both functionality and IF.
These instances with actor's solutions are fed into an LLM for adversarial analysis. The LLM is prompted to:
\begin{itemize}[leftmargin=*]
    \item \textbf{Diagnose:} Analyze why the current constraint $\tau$ was easily satisfied by the actor (e.g., identifying loopholes, loose parameter boundaries, or lack of corner-case coverage).
    \item \textbf{Mutate:} Propose an evolved schema $\tau'$ that strictly tightens the constraint or closes the identified loophole, such that the actor's original solution would fail under $\tau'$.
    \item \textbf{Validate:} Generate two positive witness examples for $\tau'$ to ensure the mutated schema remains logically sound.
\end{itemize}
Only mutated schemas that pass the validation check are incorporated into $\mathcal{T}$. 
This adversarial process effectively forces the actor to generalize rather than exploit specific problem patterns in the instruction design.

\paragraph{Discussion}
Fundamentally, the generated problem distribution is determined by the interplay between the actor model and the schema library. 
Our co-evolving paradigm optimizes this distribution by advancing these components in a mutually complementary manner, where the actor's capability growth necessitates higher complexity while the evolving library supplies the necessary structural diversity.




\section{Experiment}

In this section, we present the experimental results of proposed \model and a curated benchmark dataset based on the generated data, i.e., \bench.


\begin{table}[t]
\centering
\begin{minipage}[c]{0.58\linewidth}
    \centering
    \caption{Comparison on public IF-Coding benchmarks.}\label{tab:benchmark_impr}
    \resizebox{0.95\linewidth}{!}{
    \begin{tabular}{r|cc|cc}
        \toprule
        & \multicolumn{2}{c|}{\textbf{IFEvalCode}} & \multicolumn{2}{c}{\textbf{CodeIF}} \\
         &  $\text{Inst.}$ & $\text{Prompt}$  & $\text{Inst.}$ & $\text{Prompt}$ \\
        \midrule
        \midrule
        \multicolumn{5}{c}{\textbf{Proprietary Models}} \\
        \midrule
        \midrule
        \texttt{GPT-5.2} & 0.9391 & 0.8571 & 0.9460 & 0.6667 \\
        \texttt{Gemini-3 Pro} & 0.9271 & 0.8286 & 0.8757 & 0.4023 \\
        \texttt{Claude-4.5-Sonnet} & 0.9273 & 0.8048 & 0.8724 & 0.3879 \\
        \texttt{Claude-4.5-Opus} & 0.9249 & 0.7952 & 0.9281 & 0.5575 \\
        \texttt{Seed-1.6-thinking} & 0.9023 & 0.7238 & 0.7949 & 0.2040 \\
        \texttt{Seed-1.8-thinking} & 0.9152 & 0.7857 & 0.6764 & 0.3621 \\
        \midrule
        \midrule
        \multicolumn{5}{c}{\textbf{Open-Source Models}} \\
        \midrule
        \midrule
        \texttt{GLM-4.7} & 0.8989 & 0.7000 & 0.7163 & 0.3563 \\
        \texttt{Kimi-K2} & 0.8606 & 0.5714 & 0.8290 & 0.2931\\
        \texttt{GPT-OSS-20B} & 0.8894 & 0.7190 & 0.7090 & 0.3649 \\
        \texttt{GPT-OSS-120B} & 0.9144 & 0.7571 & 0.5910 & 0.3362 \\
        \texttt{Seed-OSS-36B} & 0.7363 & 0.6381 & 0.8231 & 0.3362 \\
        \texttt{Qwen3-Coder-480B} & 0.8466 & 0.5762 & 0.8144 & 0.2358 \\
        \texttt{DeepSeek-V3.2} & 0.8923 & 0.6741 & 0.8747 & 0.3506 \\
        \texttt{DeepSeek-Coder-V2} & 0.7681 & 0.4190 & 0.7274 & 0.1494 \\
        \midrule
        \midrule
        \multicolumn{5}{c}{\textbf{\model}} \\
        \midrule
        \midrule
        \texttt{Qwen2.5-Coder-7B} & 0.6837 & 0.2800 & 0.5986 & 0.0718 \\
        \gtext{+} 1st-\textit{Generation} & 0.8298 & 0.5857 & 0.8281 & 0.2557  \\
        \gtext{+} 2nd-\textit{Generation} & 0.8458 & 0.5904 & 0.8357 & 0.2844  \\
        \gtext{+} 3rd-\textit{Generation} & 0.8701 & 0.6143 & 0.8435 & 0.3007  \\
        \midrule
        \texttt{Seed-Coder-8B} & 0.8051 & 0.4667 & 0.7337 & 0.1523 \\
        \gtext{+} 1st-\textit{Generation} & 0.8382 & 0.5571 & 0.8430 & 0.2902 \\
        \gtext{+} 2nd-\textit{Generation} & 0.8552 & 0.5952 & 0.8509 & 0.3017 \\
        \gtext{+} 3rd-\textit{Generation} & 0.8519 & 0.6095 & 0.8520 & 0.3333 \\
        \bottomrule
    \end{tabular}
    }
\end{minipage}
\hfill
\begin{minipage}[c]{0.38\linewidth}
    \centering
    \caption{Accuracy analysis on IFEvalCode.}\label{tab:accuracy}
    \resizebox{0.9\linewidth}{!}{
    \begin{tabular}{c|cccc}
        \toprule
        Actor & Original & \gtext{+} 1st & \gtext{+} 2nd & \gtext{+} 3rd \\
        \midrule
        \midrule
        \texttt{Qwen2.5-Coder-7B} & 0.1523 & 0.2230 & 0.2523 & 0.2571 \\
        \texttt{Seed-Coder-8B} & 0.2904 & 0.2952 & 0.3285 & 0.3142 \\
        \bottomrule
    \end{tabular}
    }
    
    \vspace{0.3cm} 
    
    \caption{Performance after training on synthetic data.}\label{tab:weak_to_strong}
    \resizebox{\linewidth}{!}{
    \begin{tabular}{r|cc|cc}
        \toprule
        & \multicolumn{2}{c|}{\textbf{IFEvalCode}} & \multicolumn{2}{c}{\textbf{CodeIF}} \\
         & Inst. & Prompt & Inst. & Prompt \\
        \midrule
        \midrule
        \texttt{DeepSeek-Coder-1.3B} & 0.6446 & 0.2381 & 0.5725 & 0.0805 \\
        \gtext{w/} RL Training & 0.7822 & 0.4333 & 0.7985 & 0.2011 \\
        \midrule
        \texttt{Qwen2.5-Coder-14B} & 0.7549 & 0.4190 & 0.7571 & 0.1351 \\
        \gtext{w/} RL Training & 0.8706 & 0.6571 & 0.8762 & 0.3563 \\
        \midrule
        \texttt{Qwen2.5-Coder-32B} & 0.8492 & 0.5762 & 0.8161 & 0.2500 \\
        \gtext{w/} RL Training & 0.8962 & 0.7095 & 0.8883 & 0.4147 \\
        \bottomrule
    \end{tabular}
    }

    \vspace{0.3cm} 

    \caption{Contamination analysis.}\label{tab:contamination}
    \resizebox{\linewidth}{!}{
    \begin{tabular}{r|cc|cc}
        \toprule
        & \multicolumn{2}{c|}{\textbf{IFEvalCode}} & \multicolumn{2}{c}{\textbf{CodeIF}} \\
         & $n=5$ & $n=13$ & $n=5$ & $n=13$ \\
        \midrule
        \midrule
        \texttt{Qwen2.5-Coder-7B} & 2.14\% & 0\% & 6.13\% & 0\% \\
        \texttt{Seed-Coder-8B} & 5.06\% & 0\% & 3.02\% & 0\% \\
        \bottomrule
    \end{tabular}
    }

    \vspace{0.3cm} 

    \caption{Ablation Study on Base Model of Generator.}\label{tab:gen_base_model}
    \resizebox{\linewidth}{!}{
    \begin{tabular}{r|cc|cc}
        \toprule
        & \multicolumn{2}{c|}{\textbf{IFEvalCode}} & \multicolumn{2}{c}{\textbf{CodeIF}} \\
         & Inst. & Prompt & Inst. & Prompt \\
        \midrule
        \midrule
        \texttt{Qwen2.5-Coder-7B} & 0.6837 & 0.2800 & 0.5986 & 0.0718 \\
        \gtext{w/} \texttt{Seed-1.6} & 0.8701 & 0.6143 & 0.8435 & 0.3007 \\
        \gtext{w/} \texttt{GPT-OSS-120B} & 0.8703 & 0.6238 & 0.8408 & 0.3161 \\
        \gtext{w/} \texttt{Qwen2.5-72B} & 0.7925 & 0.4857 & 0.7753 & 0.1667 \\
        \midrule
        \texttt{Seed-Coder-8B} & 0.8051 & 0.4667 & 0.7337 & 0.1523 \\
        \gtext{w/} \texttt{Seed-1.6} & 0.8519 & 0.6095 & 0.8520 & 0.3333 \\
        \gtext{w/} \texttt{GPT-OSS-120B} & 0.8619 & 0.6333 & 0.8400 & 0.3100 \\
        \gtext{w/} \texttt{Qwen2.5-72B} & 0.8438 & 0.5857 & 0.8024 & 0.2098 \\
        \bottomrule
    \end{tabular}
    }
\end{minipage}
\end{table}

\subsection{Results of \model}\label{sec:results}

\paragraph{Source Dataset}
We clean up and construct a coding dataset by integrating three diverse public datasets, including MBPP~\cite{austin2021program}, LeetCode (Easy Level)~\cite{xia2025leetcodedataset}, ClassEval~\cite{du2023classeval}.
These datasets span distinct programming domains, covering basic logic, competitive algorithm, and class-level design.
We filter out problems with ambiguous descriptions, specifically instances where the test cases test cases are under-specified. 
Furthermore, we augment the dataset by generating missing function signatures and input/output examples. The final dataset comprises 1,565 coding problems, with 958, 519, and 88 samples derived from each respective source.

\paragraph{Benchmark}
We evaluate our method using two recent benchmarks dedicated to IF coding tasks: IFEvalCode~\cite{yang2025ifevalcode} and CodeIF~\cite{yan2025codeif}.
As our study focuses on Python, we exclusively conduct evaluations on the Python subset of these benchmarks.
Crucially, these benchmarks are curated from sources distinct from our seed dataset, ensuring no data overlap.
Using an LLM to verify instruction compliance, we report two granularities as defined in~\cite{zhou2023instruction}:
\begin{itemize}[leftmargin=*]
\item \textit{Inst.:} The average proportion of satisfied instructions per problem (Instruction-level pass rate).
\item \textit{Prompt:} The proportion of problems where \textit{all} instructions are satisfied (Prompt-level pass rate).
\end{itemize}


\paragraph{Base Model}
We evaluate the performance of 13 representative LLMs, comprising 5 proprietary models and 8 open-source reasoning models. 
Specifically, we employ \texttt{Seed-1.6} as the generator and utilize two coder models, i.e., \texttt{Qwen2.5-Coder-7B} and \texttt{Seed-Coder-8B}, as actor models, reporting their performance throughout the evolution. 
By default, we use the \texttt{Instruct} variants for all models. 
All the prompts can be found in Appendix~\ref{app:prompt}.

\paragraph{Hyperparameter \& Implementation}
We set $C=3$, $\alpha=10$, $\eta=0.5$, $\lambda=1$, and $\delta=0.1$. We pick 3 schema with the lowest values for mutation.
GRPO~\cite{shao2024deepseekmath} is applied to post-train actor models across evolving process.
As for the reward function, we define two indicator terms: $r_{\text{func}} = \mathbb{I}(f_{\text{test}}(s))$ for functional correctness and $r_{\text{IF-dense}} = \frac{1}{|\boldsymbol{\iota}|}\sum_{\iota} \mathbb{I}(f_{\text{IF-test}}^\iota(s))$ for instruction adherence rate. The final reward is computed as:
\begin{equation}\label{equ:dense_reward}
r = r_{\text{func}} + r_{\text{IF-dense}} + 2 \cdot (r_{\text{func}} \cdot r_{\text{IF-dense}})
\end{equation}
where we incentivize satisfying both constraints simultaneously.
Additional experiments of reward functions can be found in Appendix~\ref{app:exp}.


\paragraph{Benchmarking Results}
As detailed in Table~\ref{tab:benchmark_impr}, proprietary models generally outperform open-source baselines, while the \texttt{GPT-OSS} series exhibits competitive capability. 
The metrics \textit{Prompt} on CodeIF are significantly lower than those on IFEvalCode across all evaluated models. 
We attribute this gap to the higher number of instructions per problem in CodeIF, which imposes a substantially more rigorous challenge. 
Crucially, \model consistently enhances both actor models, ultimately achieving performance comparable to SOTA reasoning models. This validates the efficacy of \model's co-evolving paradigm.

In addition, Table~\ref{tab:accuracy} reports the percentage of problems satisfying both unit tests and all instruction constraints. While both models show clear gains driven by enhanced instruction adherence, \texttt{Seed-Coder-8B} exhibits more moderate improvement, as its stronger baseline leaves limited room for further growth.

We evaluate the contamination of the training dataset generated across all iterations on IFEvalCode and CodeIF, presenting the n-gram overlap results in Table~\ref{tab:contamination}. 
It can be found that the contamination rate remains at 0\% under the strict $n=13$ standard, confirming that our generated data is free from leakage and that the observed performance gains stem from genuine capability enhancement.


\begin{table*}[t]
\centering
\caption{Comparison on \bench.}\label{tab:new_benchmark}
\resizebox{1.0\linewidth}{!}{
    \begin{tabular}{r|c|cc|cc|c|cc|cc|c|cc|cc}
        \toprule
        & \multicolumn{5}{c|}{$1\sim 3$ Instructions} & \multicolumn{5}{c|}{$4\sim 6$ Instructions} & \multicolumn{5}{c}{$\ge 7$ Instructions} \\
         & \multirow{2}{*}{Func.} & \multicolumn{2}{c|}{Unit Test} & \multicolumn{2}{c|}{LLM} & \multirow{2}{*}{Func.} & \multicolumn{2}{c|}{Unit Test} & \multicolumn{2}{c|}{LLM} & \multirow{2}{*}{Func.} & \multicolumn{2}{c|}{Unit Test} & \multicolumn{2}{c}{LLM} \\
         & & Inst. & Prompt & Inst. & Prompt & & Inst. & Prompt & Inst. & Prompt & & Inst. & Prompt & Inst. & Prompt \\
        \midrule
        \midrule
        \multicolumn{16}{c}{\textbf{Proprietary Models}} \\
        \midrule
        \midrule
        \texttt{GPT-5.2} & 0.9583 & 0.9768 & 0.9652 & 0.9826 & 0.9513 & 0.9693 & 0.9800 & 0.9080 & 0.9720 & 0.8773 & 0.8720 & 0.9453 & 0.7680 & 0.9542 & 0.8160 \\
        \texttt{Gemini-3 Pro} & 0.9166 & 0.9803 & 0.9652 & 0.9953 & 0.9861 & 0.8429 & 0.9775 & 0.9157 & 0.9767 & 0.9386 & 0.7680 & 0.9602 & 0.7920 & 0.9361 & 0.7840 \\
        \texttt{Claude-4.5-Sonnet} & 0.9236 & 0.9534 & 0.9097 & 0.9710 & 0.9375 & 0.8507 & 0.9604 & 0.8352 & 0.9583 & 0.8390 & 0.7920 & 0.9618 & 0.7360 & 0.9650 & 0.7600 \\
        \texttt{Claude-4.5-Opus} & 0.9375 & 0.9806 & 0.9444 & 0.9791 & 0.9375 & 0.9003 & 0.9653 & 0.8735 & 0.9650 & 0.9003 & 0.8160 & 0.9541 & 0.7920 & 0.9254 & 0.7760 \\
        \texttt{Seed-1.6-thinking} & 0.9305 & 0.9324 & 0.8819 & 0.9675 & 0.9305 & 0.8199 & 0.9448 & 0.7892 & 0.9480 & 0.7624 & 0.6560 & 0.8840 & 0.6560 & 0.9337 & 0.6000 \\
        \texttt{Seed-1.8-thinking} & 0.9027 & 0.9571 & 0.9236 & 0.9745 & 0.9305 & 0.8084 & 0.9495 & 0.8199 & 0.9515 & 0.7892 & 0.6800 & 0.9287 & 0.7120 & 0.9516 & 0.7280 \\
        \midrule
        \midrule
        \multicolumn{16}{c}{\textbf{400B+ Open-Source Models}} \\
        \midrule
        \midrule
        \texttt{Kimi-K2} & 0.8125 & 0.8331 & 0.7291 & 0.8877 & 0.7361 & 0.7241 & 0.8684 & 0.5325 & 0.8637 & 0.5172 & 0.6880 & 0.8268 & 0.2080 & 0.8288 & 0.2240 \\
        \texttt{DeepSeek-V3.2} & 0.8680 & 0.8412 & 0.7569 & 0.9050 & 0.7708 & 0.8084 & 0.9310 & 0.6973 & 0.9141 & 0.6360 & 0.7120 & 0.9186 & 0.4880 & 0.9057 & 0.4720 \\
        \texttt{Qwen3-Coder-480B} & 0.8611 & 0.8033 & 0.6527 & 0.8761 & 0.6944 & 0.8199 & 0.8867 & 0.6015 & 0.8805 & 0.5593 & 0.7600 & 0.8830 & 0.3680 & 0.8622 & 0.3440 \\
        \midrule
        \midrule
        \multicolumn{16}{c}{\textbf{100B+ Open-Source Models}} \\
        \midrule
        \midrule
        \texttt{GLM-4.7} & 0.8819 & 0.9569 & 0.9305 & 0.9363 & 0.9097 & 0.7164 & 0.8834 & 0.7739 & 0.8371 & 0.7279 & 0.6400 & 0.9133 & 0.7360 & 0.8267 & 0.6960 \\
        \texttt{DeepSeek-Coder-V2} & 0.8194 & 0.6806 & 0.5000 & 0.7766 & 0.4861 & 0.6858 & 0.8109 & 0.3984 & 0.7763 & 0.3371 & 0.6160 & 0.7776 & 0.1280 & 0.7211 & 0.0960 \\
        \texttt{GPT-OSS-120B} & 0.9236 & 0.9623 & 0.9375 & 0.9745 & 0.9444 & 0.8659 & 0.9399 & 0.8505 & 0.9295 & 0.8314 & 0.7440 & 0.8960 & 0.7040 & 0.8477 & 0.7040 \\
        \midrule
        \midrule
        \multicolumn{16}{c}{\textbf{20B+ Open-Source Models}} \\
        \midrule
        \midrule
        \texttt{Seed-OSS-36B} & 0.8888 & 0.9108 & 0.8541 & 0.9594 & 0.8958 & 0.7816 & 0.9246 & 0.7739 & 0.9551 & 0.8314 & 0.6640 & 0.8839 & 0.6400 & 0.9660 & 0.7920 \\
        \texttt{Qwen2.5-Coder-32B} & 0.6805 & 0.8038 & 0.6805 & 0.8750 & 0.7013 & 0.5977 & 0.8658 & 0.5440 & 0.8819 & 0.5747 & 0.5600 & 0.8320 & 0.2880 & 0.8537 & 0.3040 \\
        \texttt{GPT-OSS-20B} & 0.9027 & 0.9187 & 0.8541 & 0.9375 & 0.8819 & 0.7624 & 0.8884 & 0.7432 & 0.8302 & 0.6973 & 0.6240 & 0.8301 & 0.5600 & 0.7210 & 0.5200 \\
        \midrule
        \midrule
        \multicolumn{16}{c}{\textbf{7B+ Open-Source Models}} \\
        \midrule
        \midrule
        \texttt{DeepSeek-Coder-V2-Lite} & 0.7777 & 0.4490 & 0.2013 & 0.6145 & 0.2083 & 0.7279 & 0.5722 & 0.0766 & 0.5504 & 0.0498 & 0.6400 & 0.5996 & 0.0160 & 0.5520 & 0.0160 \\
        \texttt{Qwen2.5-Coder-14B} & 0.7083& 0.6127& 0.4166& 0.7407& 0.4236& 0.6628& 0.7505& 0.2643& 0.7272& 0.2528& 0.5840& 0.6874& 0.0880& 0.6611& 0.0800 \\
        \texttt{Seed-Coder-8B} & 0.7291 & 0.6664 & 0.4652 & 0.7581 & 0.4375 & 0.6283 & 0.7789 & 0.3678 & 0.7657 & 0.3295 & 0.5920 & 0.7360 & 0.1120 & 0.7370 & 0.1200 \\
        \texttt{Qwen2.5-Coder-7B} & 0.6875 & 0.5025 & 0.2638 & 0.6469 & 0.2638 & 0.6743 & 0.6579 & 0.1417 & 0.6127 & 0.1111 & 0.5840 & 0.6425 & 0.0240 & 0.5956 & 0.0320 \\
        \bottomrule
    \end{tabular}
}
\end{table*}

\paragraph{Weak-to-Strong Scalability}
To investigate the cross-model generalizability of our synthesized data, we employ the corpus generated by the smaller actor (\texttt{Seed-Coder-8B}) to supervise larger models. 
Specifically, we aggregate the synthesized data across all three iterations to post-train three models of varying sizes, utilizing Equ.~\ref{equ:dense_reward} as the reward signal. 
The finalized dataset includes 4,241 coding problems, and more details can be found in Appendix~\ref{app:dataset}.
As presented in Table~\ref{tab:weak_to_strong}, all models exhibit significant improvements in instruction following. Most notably, \texttt{Qwen2.5-Coder-32B} achieves performance parity with proprietary SOTA models, validating the effectiveness of our weak-to-strong supervision.

\begin{figure}[t]
    \centering
    \begin{minipage}{0.5\textwidth}
        \centering
        \includegraphics[width=\textwidth]{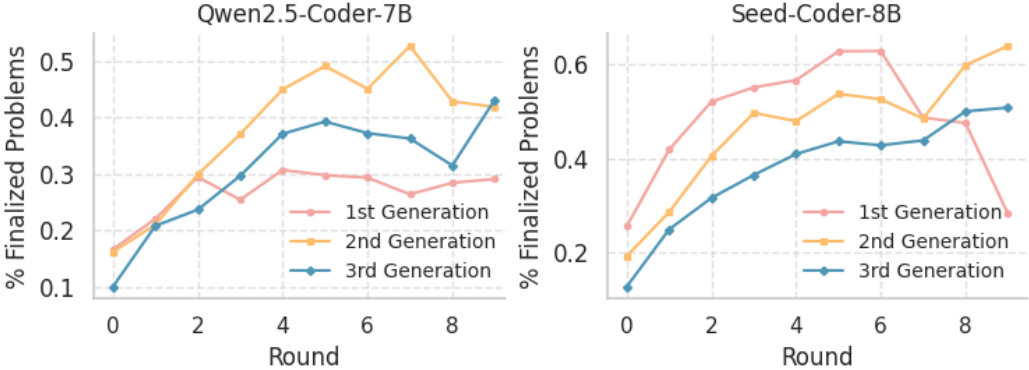}
        \caption{Percentage of finalized problems across rounds.}
        \label{fig:win_rate}
    \end{minipage}
    \hfill
    \begin{minipage}{0.45\textwidth}
        \centering
        \includegraphics[width=\textwidth]{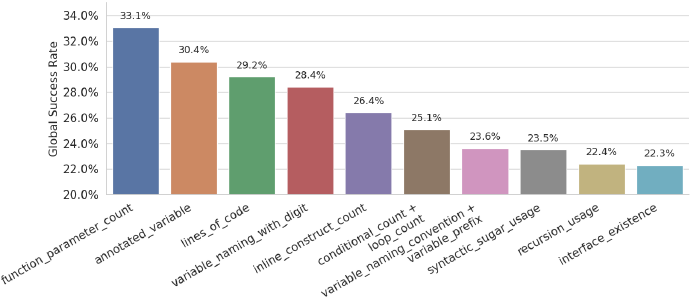}
        \caption{Schema with top 10 global success rate.}
        \label{fig:schema_win}
    \end{minipage}
    \vspace{-0.2cm}
\end{figure}



\begin{figure*}[t]
    \centering
    \includegraphics[width=1.\textwidth]{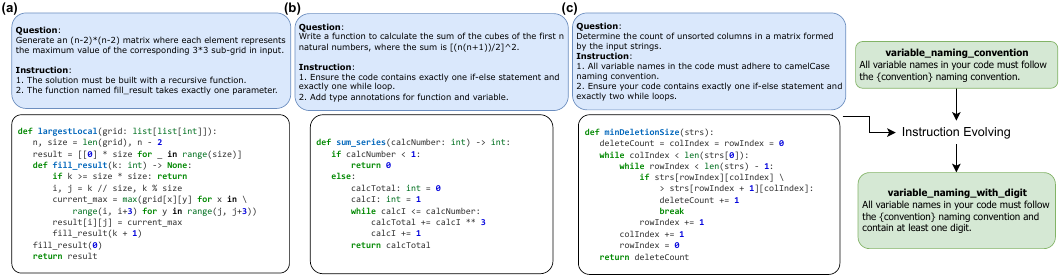}
    \caption{
        Representative examples of the three schema categories with adapted solutions. 
        \textbf{(a)} A constraint from the initial library, i.e., \lstinline{function_parameter_count} (2nd instruction). 
        \textbf{(b)} A composed schema combining control flow constraints, i.e., \lstinline{conditional_count + loop_count} (1st instruction). 
        \textbf{(c)} A mutated schema illustrating the evolution from a standard convention (\lstinline{variable_naming_convention}) to a stricter variant requiring digits (\lstinline{variable_naming_with_digit}).
    }
    \label{fig:case_study}
    \vspace{-0.2cm}
\end{figure*}

\subsection{\bench}

We curated a high-quality benchmark from our generated data using a strict human-in-the-loop verification process. 
We designed a rubric consisting of four binary sanity checks and two scalar quality metrics. 
An initial LLM-based filter first discarded invalid samples and identified high-potential candidates (Value = 3). 
These candidates then underwent manual review, where human annotators applied the same criteria. Only instances confirmed by humans to possess high value were included in the final test set with 530 problems. 
Full details are provided in Appendix~\ref{app:rubric}.

As shown in Table~\ref{tab:new_benchmark}, we divide \bench into three distinct subsets based on the number of instructions, and report five key metrics to capture both functional correctness and instruction-following capability:
\begin{itemize}[leftmargin=*]
\item \textit{Func.}(Functional Correctness): The pass rate on the original functional unit tests.
\item \textit{Inst.} and \textit{Prompt}: We apply the same metrics as mentioned in Section~\ref{sec:results} with either rigorous execution-based checkers or LLM-based judge.
\end{itemize}
The evaluation results derived from the LLM judge align closely with those from the execution-based Unit Tests, demonstrating the reliability and accuracy of our execution-based verification.
Besides, a clear inverse correlation between the number of instructions and performance can be observed. 
A substantial performance gap exists between proprietary models and smaller open-source models, where smaller models struggle significantly to satisfy complex, concurrent constraints.

\subsection{Further Analysis}

\paragraph{Sampling Efficiency} 
To analyze sampling efficiency, we track the ratio of finalized problems to total sampled candidates per round. 
Specifically, the actor's pass rate serves a routing signal, where a problem is deemed finalized if the actor fails to solve the augmented problem (Section~\ref{sec:inner_loop}).
An efficient sampler is expected to learn from interaction, progressively shifting the sampling distribution towards hard-to-solve cases.
As visualized in Figure~\ref{fig:win_rate}, the finalization rate initially climbs and converges as instructions accumulate, demonstrating search efficiency. 
Besides, the two actor models exhibit distinct learning dynamics. For the stronger \texttt{Seed-Coder-8B}, the sampling process becomes increasingly challenging after evolution.

\paragraph{Case Study} 
Figure~\ref{fig:schema_win} illustrates the top-10 instruction schema ranked by global success rate, i.e., $P(\pi)$, for the 3rd-generation \texttt{Seed-Coder-8B}.
It can be found that top-ranked schema comprise initial constraints (e.g., \lstinline{function_parameter_count}), composite constraints (e.g., \lstinline{conditional_count+loop_count}), and mutated variants (e.g., \lstinline{variable_naming_with_digit}). 
The detailed definition can be found in Appendix~\ref{app:dataset}.

Furthermore, we showcase representative examples for each of these three categories in Figure~\ref{fig:case_study}.
The case study highlights three distinct difficulty mechanisms. 
Case (a) shows that even simple constraints like \lstinline{function_parameter_count} become challenging when coupled with contextual requirements such as recursion. 
Case (b) showcases the complexity arising from the composition of multiple control-flow constraints. 
Case (c) exemplifies the adversarial evolution process: the sampler detects patterns in the current solution and synthesizes a mutated constraint, i.e., enforcing digits in variable names, that renders the previous solution invalid.

\paragraph{Ablation Study on $\pi_{\text{gen}}$}
We analyze the effect of the generator backbone $\pi_{\text{gen}}$.
In addition to our default \texttt{Seed-1.6}, we test two open-source alternatives, i.e., \texttt{GPT-OSS-120B} (reasoning model) and \texttt{Qwen2.5-72B} (non-reasoning IF model). 
Table~\ref{tab:gen_base_model} demonstrates that our method drives consistent improvements across all generator configurations. 
Importantly, the open-source \texttt{GPT-OSS-120B} achieves parity with the \texttt{Seed-1.6} model, underscoring the feasibility and robustness of our approach.
Ablation study on the core modules can be found in Appendix~\ref{app:exp}.


\section{Conclusion}
We presented \model, a co-evolutionary data synthesis framework that unifies parametric instruction design with MCTS-guided validation. By structurally evolving the schema library alongside the actor model, our approach guarantees strict solvability while continuously pushing the frontier of problem complexity. Empirical results demonstrate that \model effectively overcomes the limitations of static data generation, establishing a robust foundation for training next-generation instruction-following code LLMs.


\clearpage

\bibliographystyle{plainnat}
\bibliography{main}

\clearpage

\beginappendix

\section{Limitation}\label{app:limitation}

A limitation of our current study is its focus on Python, chosen for its dominance in algorithmic reasoning benchmarks and competitive programming. 
While the proposed co-evolutionary framework and MCTS sampler are inherently language-agnostic, extending \model to a multi-lingual setting (e.g., C++, Java, or Rust) requires adapting the underlying AST verifiers and defining language-specific instruction schemas. 
Future work will focus on generalizing the schema library to support diverse programming syntaxes and broader software engineering paradigms beyond algorithmic tasks.

\section{Implementation}\label{app:experimental_setup}

\paragraph{Pseudocode} Here we provide the pesudocode of our proposed \model in Algo.\ref{algo:framework}.
Besides, we apply a dynamic expansion rule: if the candidate pool size drops below 500 (due to routing), each problem is augmented twice in the next iteration. This redundant augmentation ensures a consistent data supply.
For each finalized question, we will apply an LLM to rephrase the instruction without modifying the logic to improve the diversity.
To enhance diversity, we employ a LLM to perform semantics-preserving paraphrasing for instructions of finalized questions. The prompt can be found in Appendix~\ref{app:prompt}.

\begin{algorithm}[h]
\caption{\model}
\label{algo:framework}
\begin{algorithmic}[1]
\REQUIRE Seed Dataset $\mathcal{D}_{\text{seed}}$, Instruction Schema Library $\mathcal{T}$, Generator $\pi_{\text{gen}}$, Actor $\pi_{\text{actor}}$

\STATE Initialize sampler $\pi_{\text{MCTS}}$ using $\mathcal{T}$
\STATE $\mathcal{D}_{\text{final}} \leftarrow \emptyset$

\FOR{iteration $k = 1$ to $K$}
    \STATE \textcolor{blue}{\textit{// Phase 1: Multi-Round Augmentation}}
    \STATE $\mathcal{D}_{\text{new}} \leftarrow \text{AugmentData}(\mathcal{D}_{\text{seed}}, \mathcal{T}, \pi_{\text{gen}}, \pi_{\text{actor}}, \pi_{\text{MCTS}})$
    
    \STATE $\mathcal{D}_{\text{final}} \leftarrow \mathcal{D}_{\text{final}} \cup \mathcal{D}_{\text{new}}$
    
    \STATE \textcolor{blue}{\textit{// Phase 2: Actor Evolving}}
    \STATE $\pi_{\text{actor}} \leftarrow \text{PostTrain}(\pi_{\text{actor}}, \mathcal{D}_{\text{new}})$

    \STATE \textcolor{blue}{\textit{// Phase 3: Sampler Evolving}}
    \STATE $\pi_{\text{MCTS}} \leftarrow \text{Reset}(\mathcal{D}_{\text{new}}, \pi_{\text{MCTS}}, \pi_{\text{actor}})$

    \STATE \textcolor{blue}{\textit{// Phase 4: Schema Evolving}}
    \STATE $\{\tau_{i \oplus j}\} \leftarrow \text{Composition}(\pi_{\text{MCTS}}, \pi_{\text{actor}})$ \COMMENT{Eq.~\ref{equ:composition_1},\ref{equ:composition_2}}
    \STATE $\mathcal{T} \leftarrow \mathcal{T} \cup \{\tau_{i \oplus j}\}$
    \STATE $\mathcal{T}_{\text{weak}} \leftarrow \text{SelectWeakSchema}(\mathcal{T}, \pi_{\text{MCTS}})$
    \STATE $\mathcal{T}_{\text{weak}}' \leftarrow \text{Mutation}(\mathcal{D}_{\text{final}},\mathcal{T}_{\text{weak}}, \pi_{\text{gen}})$
    \STATE $\mathcal{T} \leftarrow \mathcal{T}/\mathcal{T}_{\text{weak}} \cup \mathcal{T}_{\text{weak}}'$
    
\ENDFOR

\RETURN $\mathcal{D}_{\text{final}}, \pi_{\text{actor}}$
\end{algorithmic}
\end{algorithm}

\paragraph{Running environment.}
The experiments are conducted on a single Linux server with Intel(R) Xeon(R) Platinum 8336C CPU, 1.9Ti RAM, and 8 NVIDIA A800-SXM4-80GB. Our method is implemented on PyTorch 2.8.0 and Python 3.11.2.

\paragraph{Training}
All the RL training are based on an open-source framework verl\footnote{\url{https://github.com/verl-project/verl}}.
We set the group size to $G=8$, with a global training batch size of 256 and a mini-batch size of 64. 
We employ dynamic batching with a maximum of 16,384 tokens per GPU.
The learning rate is fixed at $1 \times 10^{-6}$ with 10 warmup steps. We employ an asymmetric PPO clipping range of $[0.2, 0.28]$ and disable the KL penalty. All training is conducted with FSDP2 and vLLM integration, supporting a maximum sequence length of 2048 tokens.

\section{Dataset}\label{app:dataset}

\subsection{Schema}

The library of parametric schema is detailed in Table~\ref{tab:schema_full}. 
For improved presentation, certain schema identifiers have been abbreviated (e.g., \texttt{variable\_existence} is denoted as \texttt{var\_existence}).

\begin{table}[]
\centering
\small
\caption{Library of parametric instruction schema.}
\label{tab:schema_full}
\begin{tabularx}{\textwidth}{l X X}
\toprule
\textbf{Schema} & \textbf{Instruction Template} & \textbf{Parameter} \\ \midrule

\rowcolor[gray]{0.9} \multicolumn{3}{l}{\textbf{Category: Variable \& Structures}} \\
\texttt{var\_existence} & Your code \{\texttt{mode}\} define a variable named \{\texttt{name}\}. & \texttt{mode}: [should, should not]; \texttt{name}: \textit{str} \\
\texttt{no\_inter\_var} & Your code must not use intermediate variables. & - \\
\texttt{naming\_conv} & Variable names must follow \{\texttt{style}\} convention. & \texttt{style}: [camelCase, PascalCase, snake\_case] \\
\texttt{global\_var} & Must define a global variable named \{\texttt{name}\}. & \texttt{name}: \textit{str} \\
\texttt{init\_value} & Define variable \{\texttt{name}\} initialized with \{\texttt{val}\}. & \texttt{name}: \textit{str}; \texttt{val}: \textit{str} \\
\texttt{name\_len} & Variable name length \{\texttt{mode}\} exceed \{\texttt{n}\} characters. & \texttt{mode}: [should, should not]; \texttt{n}: \textit{int} \\
\texttt{var\_prefix} & Variable names must start with prefix `\{\texttt{pre}\}'. & \texttt{pre}: \textit{str} \\
\texttt{var\_suffix} & Variable names must end with suffix `\{\texttt{suf}\}'. & \texttt{suf}: \textit{str} \\ 
\texttt{data\_struct} & The data structure \{\texttt{ds}\} \{\texttt{mode}\} be used. & \texttt{ds}: [dict, set, list, tuple, ...]; \texttt{mode}: [must, must not] \\
\midrule

\rowcolor[gray]{0.9} \multicolumn{3}{l}{\textbf{Category: Logic \& Control Flow}} \\
\texttt{loop\_count} & Include \{\texttt{comp}\} \{\texttt{n}\} \{\texttt{type}\} loop(s). & \texttt{comp}: [at least, at most, exactly]; \texttt{type}: [while, for]; \texttt{n}: \textit{int} \\

\texttt{forbid\_loop} & Your code must not use any \{\texttt{type}\} loops. & \texttt{type}: [while, for] \\

\texttt{cond\_count} & Include \{\texttt{comp}\} \{\texttt{n}\} if-else statement(s). & \texttt{comp}: [exactly, at least, at most]; \texttt{n}: \textit{int} \\
\texttt{switch\_stmt} & Must include a switch (or match/case) statement. & - \\
\texttt{recursion} & You must implement the solution using recursion. & - \\

\midrule

\rowcolor[gray]{0.9} \multicolumn{3}{l}{\textbf{Category: Interface \& Type}} \\
\texttt{interface} & You should define a \{\texttt{type}\} named \{\texttt{name}\}. & \texttt{type}: [class, interface]; \texttt{name}: \textit{str} \\
\texttt{type\_hint} & Type annotations must be used for functions and vars. & - \\
\texttt{func\_def} & Your code must define a function named \{\texttt{name}\}. & \texttt{name}: \textit{str} \\
\texttt{param\_count} & Function \{\texttt{name}\} must accept exactly \{\texttt{n}\} params. & \texttt{name}: \textit{str}; \texttt{n}: \textit{int} \\

 \midrule

\rowcolor[gray]{0.9} \multicolumn{3}{l}{\textbf{Category: Library \& Tools}} \\
\texttt{import\_lib} & Your code \{\texttt{mode}\} import the library \{\texttt{lib}\}. & \texttt{mode}: [must, must not]; \texttt{lib}: \textit{str} \\
\texttt{no\_import} & Your code must not import any library. & - \\
\texttt{forbid\_func} & You are not allowed to use built-in(s): \{\texttt{funcs}\}. & \texttt{funcs}: \textit{list[str]} \\
\texttt{require\_func} & You must use the built-in function(s): \{\texttt{funcs}\}. & \texttt{funcs}: \textit{list[str]} \\
\texttt{sugar\_usage} & Your code \{\texttt{mode}\} utilize \{\texttt{type}\}. & \texttt{mode}: [must, must not]; \texttt{type}: [list comp, lambda, ternary operator, generator exp, ...] \\
\texttt{sugar\_count} & Utilize \{\texttt{comp}\} \{\texttt{n}\} \{\texttt{type}\}. & \texttt{type}: [list comp, lambda, ternary operator, generator exp, ...]; \texttt{n}: \textit{int}; \texttt{comp}: [at least, at most, exactly] \\ 

\midrule
\rowcolor[gray]{0.9} \multicolumn{3}{l}{\textbf{Category: Style \& Formatting}} \\
\texttt{comment\_cnt} & Include \{\texttt{comp}\} \{\texttt{n}\} line(s) of comments. & \texttt{comp}: [exactly, more than, less than]; \texttt{n}: \textit{int} \\
\texttt{comment\_lang}& Comments must be written in \{\texttt{lang}\}. & \texttt{lang}: [en, zh] \\

\texttt{code\_lines} & The solution must contain \{\texttt{comp}\} \{\texttt{n}\} lines of code. & \texttt{comp}: [exactly, at least, at most]; \texttt{n}: \textit{int} \\

\bottomrule
\end{tabularx}
\end{table}

Each schema is paired with an AST-based verification function, which dynamically adapts to the instantiated parameters. 
Taking the \texttt{variable\_existence} (Code~\ref{code:ast_verify}) and \texttt{conditional\_count} (Code~\ref{code:ast_verify_cond_count}) as examples, the verifier traverses the Abstract Syntax Tree to identify all variable assignments and control flow structures. 

In \texttt{variable\_existence}, the parameter \texttt{mode} governs the validation logic: if set to `should', the function confirms the presence of the specified variable; if 'should not', it ensures its absence. As for \texttt{conditional\_count}, the parameter \texttt{comparison} (e.g., `at least', `exactly') combined with a target \texttt{count} defines the numerical constraint. The verifier counts the occurrences of \texttt{ast.If} and \texttt{ast.Compare} nodes, subsequently evaluating whether the total satisfies the specified relational condition.

\begin{listing}[h]
    \begin{lstlisting}[style=python, frame=none]
import ast

def check_variable_existence(code, variable_name, mode):
    tree = ast.parse(code)
    assert mode in ["should", "should not"]
    defined_vars = set()
    for node in ast.walk(tree):
        if isinstance(node, ast.Assign):
            for target in node.targets:
                if isinstance(target, ast.Name):
                    defined_vars.add(target.id)
        # Also catch annotated assignment: x: int = 1
        elif isinstance(node, ast.AnnAssign):
            if isinstance(node.target, ast.Name):
                defined_vars.add(node.target.id)
    exists = variable_name in defined_vars
    if str(mode).lower() == "should not":
        return not exists
    return exists
    \end{lstlisting}
    \caption{AST-based verification logic for \texttt{variable\_existence}.}
    \label{code:ast_verify}
\end{listing}

\begin{listing}[h]
    \begin{lstlisting}[style=python, frame=none]
import ast

def check_conditional_count(code, count, comparison):
    tree = ast.parse(code)
    assert comparison in ["at least", "at most", "exactly"]
    if_nodes = [n for n in ast.walk(tree) if isinstance(n, ast.If)]
    actual = len(if_nodes)
    target = int(count)
    if "at least" in comparison: return actual >= target
    if "at most" in comparison: return actual <= target
    if "exactly" in comparison: return actual == target
    return False
    \end{lstlisting}
    \caption{AST-based verification logic for \texttt{conditional\_count}.}
    \label{code:ast_verify_cond_count}
\end{listing}





\newpage
\subsection{Training Dataset}

\begin{wrapfigure}{r}{0.40\textwidth}
    \centering
    \includegraphics[width=0.35\textwidth]{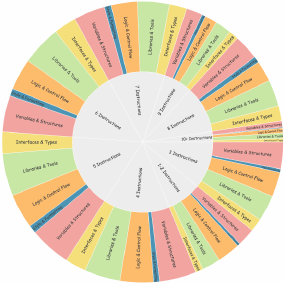}
    \caption{The distributions of schema category and the number of instructions.}
    \label{fig:dataset}
\end{wrapfigure}

In Section~\ref{sec:results}, we evaluate the efficacy of our approach by post-training a series of LLMs across different parameter scales using the synthesized dataset. 
The dataset was curated via \model, employing \texttt{Seed-Coder-8B} as the base actor model. 
Our pipeline yielded a total of 4,241 high-quality IF coding instances, with an average of approximately 1,400 problems generated per iteration. The taxonomic distribution of schema categories and the corresponding instruction counts are detailed in Figure~\ref{fig:dataset}.

It can be found that our dataset exhibits a robust hierarchy of complexity. 
The inner circle denotes the number of concurrent constraints per problem, ranging from 1 to over 10. Notably, a significant portion of the dataset comprises problems with 4 to 9 instructions, indicating a high level of task difficulty. 
The outer ring reveals a balanced distribution across functional domains, such as \textit{Logic \& Control Flow} and \textit{Interfaces \& Types}, ensuring that the model undergoes comprehensive training across diverse coding dimensions.

\paragraph{Examples}
We present representative examples of the generated IF coding data across varying complexity levels, specifically featuring 2, 4, 6, and 8 instructions. 
Each instance comprises a formal problem definition, a list of constraints, example input-output pairs, a function signature, and a reference solution. 
Note that the foundational instruction, i.e., \textit{'Implement the solution in Python'}, is treated as a global prerequisite and is excluded from the total instruction count.

\begin{PromptTealBox}{Example with 2 instructions}
\# Problem Description
\newline
Write a function that takes a list of tuples where each tuple contains a name and a mark. The function should compute the total marks for each name by summing their marks, then return the tuple consisting of the name with the highest total marks and their total. 
\newline
\newline
\# Instruction
\newline
1. Implement the solution using Python.
\newline
2. It is forbidden to use the built-in 'max' function in your code.
\newline
3. The code must make use of list comprehension. 
\newline
\newline
\# Example Input and Output
\newline
Example Input 1: [('Juan Whelan',90),('Sabah Colley',88),('Peter Nichols',7),('Juan Whelan',122),('Sabah Colley',84)]
\newline
Example Output 1: ('Juan Whelan', 212)
\newline
Example Input 2: [('Juan Whelan',10),('Sabah Colley',20),('Peter Nichols',30),('Juan Whelan',40),('Sabah Colley',50)]
\newline
Example Output 2: ('Sabah Colley', 70) 
\newline
\newline
\# Function Signature
\begin{lstlisting}[style=python, frame=none]
def max_aggregate(stdata):
    pass
\end{lstlisting}
\# Solution
\begin{lstlisting}[style=python, frame=none]
def max_aggregate(stdata):
    aggregated = [
        (name, sum(mark for n, mark in stdata if n == name)) \
            for name in {n for n, _ in stdata}
    ]
    max_name = None
    max_total = 0
    for name, total in aggregated:
        if max_name is None or total > max_total:
            max_name = name
            max_total = total
    return (max_name, max_total) 
\end{lstlisting}
\end{PromptTealBox}

\begin{PromptTealBox}{Example with 4 instructions}
\# Problem Description
\newline
Write a Python function to check whether the given number can be represented as the sum of two squares of positive integers. 
\newline
\newline
\# Instruction
\newline
1. Develop the solution using Python.
\newline
2. Define a variable called found and set its initial value to False in your code.
\newline
3. Utilize the set data structure in your implementation.
\newline
4. Include a switch (or match/case) statement within the code.
\newline
5. Make sure to import the math library in your code. 
\newline
\newline
\# Example Input and Output
\newline
Example 1:
\newline
Input: 25
\newline
Output: True (since 3² + 4² = 25)
\newline
Example 2:
\newline
Input: 24
\newline
Output: False 
\newline
\newline
\# Function Signature
\begin{lstlisting}[style=python, frame=none]
def sum_Square(n):
    pass 
\end{lstlisting}
\# Solution
\begin{lstlisting}[style=python, frame=none]
import math

def sum_Square(n):
    found = False
    squares = set()
    i = 1
    while i <= math.isqrt(n):
        square = i * i
        squares.add(square)
        remainder = n - square
        match remainder:
            case x if x in squares and x != 0:
                found = True
                break
        i += 1
    return found 
\end{lstlisting}
\end{PromptTealBox}

\begin{PromptTealBox}{Example with 6 instructions}
\# Problem Description
\newline
Write a function to determine if a given number is abundant. An abundant number is a number for which the sum of its proper divisors (excluding the number itself) is greater than the number. 
\newline
\newline
\# Instruction
\newline
1. Implement the solution using Python.
\newline
2. Ensure every variable name in your code adheres to the snake\_case naming convention.
\newline
3. Do not import the math library in your code.
\newline
4. Make use of list comprehension in your implementation.
\newline
5. Avoid using any for loops in your code.
\newline
6. Incorporate the set data structure in your solution.
\newline
7. Include exactly one while loop in your code. 
\newline
\newline
\# Example Input and Output
\newline
Example Input 1: 12
\newline
Example Output 1: True
\newline
Example Input 2: 15
\newline
Example Output 2: False 
\newline
\newline
\# Function Signature
\begin{lstlisting}[style=python, frame=none]
def check_abundant(n):
    pass
\end{lstlisting}
\# Solution
\begin{lstlisting}[style=python, frame=none]
def check_abundant(n):
    divisors = set()
    i = 1
    while i * i <= n:
        if n % i == 0:
            divisors.add(i)
            if i != n // i:
                divisors.add(n // i)
        i += 1
    proper_sum = sum([d for d in divisors if d != n])
    return proper_sum > n
\end{lstlisting}
\end{PromptTealBox}

\begin{PromptTealBox}{Example with 8 instructions}
\# Problem Description
\newline
Given a list of toy types, return true if a permutation of the list could form a palindrome sequence (reads the same forwards and backwards) and false otherwise. 
\newline
\newline
\# Instruction
\newline
1. Implement the solution using Python.
\newline
2. The code must include exactly one list comprehension and one while loop.
\newline
3. Ensure your code does not define a variable with the name "index".
\newline
4. It is mandatory to incorporate a switch (or match/case) statement in the code.
\newline
5. The solution must make use of a generator expression.
\newline
6. You must employ the built-in `collections.Counter` function.
\newline
7. A global variable named `palindrome\_check\_enabled` must be defined in the code.
\newline
8. Full type annotations (type hints) are required for every function and variable.
\newline
9. The built-in `sum` function is strictly prohibited from use in the solution. 
\newline
\newline
\# Example Input and Output
\newline
Example 1:
\newline
Input: s = "code"
\newline
Output: false
\newline
\newline
Example 2:
\newline
Input: s = "aab"
\newline
Output: true
\newline
\newline
Example 3:
\newline
Input: s = "carerac"
\newline
Output: true 
\newline
\newline
\# Function Signature
\begin{lstlisting}[style=python, frame=none]
def canPermutePalindrome(s: str) -> bool:
    pass 
\end{lstlisting}
\# Solution
\begin{lstlisting}[style=python, frame=none]
from collections import Counter
from typing import Generator, Iterable, Iterator

palindrome_check_enabled: bool = True

def canPermutePalindrome(s: str) -> bool:
    global palindrome_check_enabled
    if not palindrome_check_enabled:
        return False
    chars: list[str] = [c for c in s]
    char_count: Counter[str] = Counter(chars) 
    odd_counts: Generator[int, None, None] = (v % 2 for v in char_count.values())
    odd_count_total: int = 0
    odd_iter: Iterator[int] = iter(odd_counts)
    while True:
        try:
            val: int = next(odd_iter)
            match val:
                case 1:
                    odd_count_total += 1
                case _:
                    pass
        except StopIteration:
            break
    return odd_count_total < 2
\end{lstlisting}
\end{PromptTealBox}

\section{Additional Experiments}\label{app:exp}

\begin{figure*}[h]
    \centering
    \includegraphics[width=1.0\textwidth]{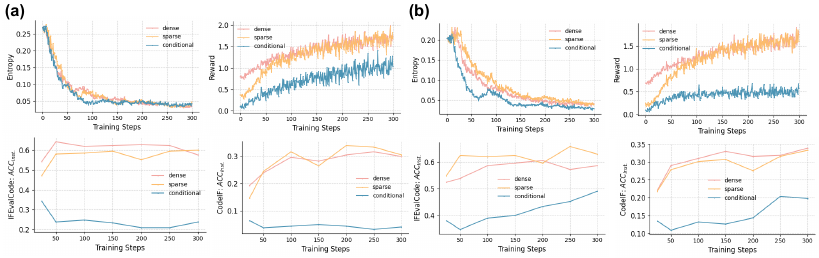}
    \caption{
        Learning curves of different RL reward functions using \texttt{Qwen2.5-Coder-7B} \textbf{(a)} and \texttt{Seed-Coder-8B} \textbf{(b)} as base models
    }
    \label{fig:post-train}
    \vspace{-0.2cm}
\end{figure*}

\paragraph{Reward Function}
We investigate the impact of reward shaping by ablating the reward function design on the same training corpus, as shown in Figure~\ref{fig:post-train}(a). Compared to the dense reward (Eq.~\ref{equ:dense_reward}), we examine two distinct variants:
\begin{itemize}[leftmargin=*] 
\item Sparse Reward: We remove the dense shaping term, providing instruction feedback only upon perfect compliance. \begin{equation} 
r = r_{\text{func}} + \mathbb{I}(r_{\text{IF-dense}} = 1) + 2 \cdot (r_{\text{func}} \cdot r_{\text{IF-dense}})
\end{equation} 

\item Conditional Reward: We employ a gated formulation where instruction adherence is rewarded only if the code is functionally correct.
\begin{equation}
    r = r_{\text{func}} \cdot \left( 1 + r_{\text{dense}} + 2.0 \cdot \mathbb{I}(r_{\text{dense}}=1) \right)
\end{equation}
\end{itemize}
As shown in Figure~\ref{fig:post-train}, we present the learning curves on entropy and reward across the training on \texttt{Seed-Coder-8B} (a) and \texttt{Qwen2.5-Coder-7B} (b).
The Conditional Reward exhibits the poorest performance due to its strict gating mechanism, which masks instruction adherence signals whenever functional correctness fails. 
This creates reward sparsity during early training. 
In contrast, the Dense/Sparse Rewards provides continuous feedback independent of functional success, facilitating a smoother optimization process.

\begin{wraptable}{r}{0.5\textwidth}
    \centering
    \vspace{-10pt}
    \caption{Ablation study on core modules.}
    \label{tab:ablation}
    \resizebox{\linewidth}{!}{ 
    \begin{tabular}{r|cc|cc}
        \toprule
        & \multicolumn{2}{c|}{\textbf{IFEvalCode}} & \multicolumn{2}{c}{\textbf{CodeIF}} \\
         & Inst. & Prompt & Inst. & Prompt \\
        \midrule
        \midrule
        \texttt{Qwen2.5-Coder-7B} & \multirow{2}{*}{0.8701} & \multirow{2}{*}{0.6143} & \multirow{2}{*}{0.8271} & \multirow{2}{*}{0.2672} \\
        \gtext{+} \model & & &  & \\
        \gtext{w/o} MCTS  & 0.8542 & 0.5952 & 0.8388 & 0.2989 \\
        \gtext{w/o} Actor Evolving  & 0.8701 & 0.6143 & 0.8435 & 0.3007 \\
        \gtext{w/o} Instruction Evolving & 0.8700 & 0.6001 & 0.8421 & 0.3132 \\
        \midrule
        \texttt{Seed-Coder-8B} & \multirow{2}{*}{0.8519} & \multirow{2}{*}{0.6095} & \multirow{2}{*}{0.8520} & \multirow{2}{*}{0.3333} \\
        \gtext{+} \model & & &  & \\
        \gtext{w/o} MCTS  & 0.8514 & 0.5905 & 0.8382 & 0.2902 \\
        \gtext{w/o} Actor Evolving  & 0.8433 & 0.5870 & 0.8421 & 0.2903 \\
        \gtext{w/o} Instruction Evolving  & 0.8521 & 0.6143 & 0.8376 & 0.2874 \\
        \bottomrule
    \end{tabular}
    }
    \vspace{-10pt} 
\end{wraptable}


\paragraph{Ablation Study} 
We conduct an ablation study to assess the effectiveness of the core components of \model.
Specifically, we consider three variants: 
(1) w/o MCTS, where the MCTS-guided search is replaced by uniform random sampling; 
(2) w/o Actor Evolving, which freezes the actor model and generates the full volume of data in a single-round synthesis; and 
(3) w/o Instruction Evolving, where the schema library remains the same.
The results presented in Table~\ref{tab:ablation} indicate that removing either core module leads to a decline in actor model's performance mostly.

\section{Rubric}\label{app:rubric}
We detail the evaluation rubric employed to curate \bench, comprising six distinct criteria. 
These are divided into two categories: Sanity Metrics and Quality Metrics. The first four, i.e., \textit{Consistency, Redundancy, Validity,} and \textit{Alignment}, serve as binary filters; any candidate failing a single criterion (scoring 0) is discarded. The remaining two, including \textit{Transformation} and \textit{Value}, assess the substantive difference between the vanilla and constrained solutions.
Our curation pipeline begins with an automated phase where \texttt{Seed-1.6} filters candidates, retaining only those meeting the sanity requirements alongside thresholds of $\text{Transformation} \ge 2$ and $\text{Value} \ge 2$. These candidates then undergo rigorous manual inspection using the same rubric. Ultimately, \bench consists of 530 gold-standard instances that satisfy all four sanity checks and achieve a maximum score of $\text{Value}=3$. The scoring strategies and representative examples for each criterion are provided below.

\begin{itemize}[leftmargin=*]
\item Consistency: Each instruction must not conflict with the problem description or other instructions.
\begin{itemize} 
    \item Scoring: 0 / 1
    \item Negative Example 1: "Ensure every variable name in your code adheres to the camelCase naming convention." AND "The code must avoid using any intermediate or temporary variables.".
    \item Negative Example 2: "Import the `math` library in your code." AND "Do not use the `math.pow` built-in function".
\end{itemize}

\item Redundancy: The instructions must be unique and necessary.
\begin{itemize} 
    \item Scoring: 0 / 1
    \item Negative Example 1: "All variable names must start with the prefix `num' and adhere to the camelCase naming convention." AND "All variable names must follow camelCase and include at least one digit.".
    \item Negative Example 2: "The code must include type annotations for every function and variable, and you are required to use the built-in `collections.Counter' function." AND "Ensure the code has exactly two lines of comments and provides type hints for all functions and variables.".
\end{itemize}

\item Validity: Instructions must be verifiable.
\begin{itemize} 
    \item Scoring: 0 / 1
    \item Negative Example 1: "The code must be elegant.".
    \item Negative Example 2: "The code must be highly readable and use meaningful variable names.".
\end{itemize}

\item Alignment: The provided solution must satisfy all instructions.
\begin{itemize} 
    \item Scoring: 0 / 1
    \item Negative Example 1: "Implement a function to calculate the factorial of $n$ using recursion" AND solution: 
\begin{lstlisting}[style=python, frame=none]
def factorial(n):
    result = 1
    for i in range(1, n + 1):
        result *= i
    return result
\end{lstlisting}

    \item Negative Example 2: "The code cannot use any built-in sorting functions" AND solution:
\begin{lstlisting}[style=python, frame=none]
def sort_array(arr):
    return sorted(arr)
\end{lstlisting}
\end{itemize}

\item Transformation: Evaluate the extent of structural modification between the vanilla solution (generated without specific constraints) and the constrained solution.
\begin{itemize} 
    \item Scoring: 1 / 2 / 3 (1: Surface, 2: Medium, 3: High)
    \item Surface-level: Cosmetic changes only. The vanilla solution: 
\begin{lstlisting}[style=python, frame=none]
import heapq
def merge_sorted_list(num1,num2,num3):
  num1 = sorted(num1)
  num2 = sorted(num2)
  num3 = sorted(num3)
  result = heapq.merge(num1,num2,num3)
  return list(result) 
\end{lstlisting}
        The constrained solution:
\begin{lstlisting}[style=python, frame=none]
import heapq
def merge_sorted_list(num1, num2, num3):
    sorted_num1 = sorted(num1)
    sorted_num2 = sorted(num2)
    sorted_num3 = sorted(num3)
    sorted_result = heapq.merge(sorted_num1, sorted_num2, sorted_num3)
    return list(sorted_result) 
\end{lstlisting}
    
    \item Medium-level: Apply different data structure, control flow, but still the same logical flow. The vanilla solution: 
\begin{lstlisting}[style=python, frame=none]
def text_match_three(text):
    for i in range(len(text) - 3):
        if text[i] == 'a' and text[i+1] == 'b' and text[i+2] == 'b' and text[i+3] == 'b':
            return 'Found a match!'
    return 'Not matched!' 
\end{lstlisting}
    The constrained solution:
\begin{lstlisting}[style=python, frame=none]
import re
def text_match_three(text):
    patterns = 'ab{3}?'
    if re.search(patterns,  text):
        return 'Found a match!'
    else:
        return('Not matched!') 
\end{lstlisting}
    \item High-level: Paradigm Shift / Algorithmic Change. The vanilla solution: 
\begin{lstlisting}[style=python, frame=none]
def permutation_coefficient(n, k): 
    P = [[0 for i in range(k + 1)] for j in range(n + 1)] 
    for i in range(n + 1): 
        for j in range(min(i, k) + 1): 
            if (j == 0): 
                P[i][j] = 1
            else: 
                P[i][j] = P[i - 1][j] + (j * P[i - 1][j - 1]) 
            if (j < k): 
                P[i][j + 1] = 0
    return P[n][k] 
\end{lstlisting}
The constrained solution:
\begin{lstlisting}[style=python, frame=none]
from functools import reduce
def permutation_coefficient(N, K):
    if K == 0:
        return 1
    if K > N:
        return 0
    return reduce(lambda X, Y: X * Y, range(N, N - K, -1)) 
\end{lstlisting}
\end{itemize}

\item Value: The goal is to assess whether the instruction leads to meaningful logical changes, algorithmic optimizations, or paradigm shifts in the code, rather than simply making changes for the sake of change (or making things worse).
\begin{itemize} 
    \item Scoring: 1 / 2 / 3 (1: Negative, 2: Neutral, 3: Positive)
    \item Negative value: Instruction forces the model to generate code that is more difficult to read, less secure, and technically unsound. The vanilla solution: 
\begin{lstlisting}[style=python, frame=none]
def extract_string(str, l):
    result = [e for e in str if len(e) == l] 
    return result 
\end{lstlisting}
The constrained solution:
\begin{lstlisting}[style=python, frame=none]
def data_Recursive_Len(data_String):
    if not data_String:
        return 0
    return 1 + data_Recursive_Len(data_String[1:])

data_Filter_Check = lambda data_String, data_Len: data_Recursive_Len(data_String) == data_Len

def extract_string(data_InputList, data_TargetLength):
    data_FilteredResult = list(
        e for e in data_InputList if data_Filter_Check(e, data_TargetLength)
    )
    return data_FilteredResult 
\end{lstlisting}

    \item Neutral value: The instructions led to changes in the code's structure, but the core logic, complexity, and engineering quality remained essentially unchanged. The vanilla solution: 
\begin{lstlisting}[style=python, frame=none]
def extract_index_list(l1, l2, l3):
    result = []
    for m, n, o in zip(l1, l2, l3):
        if (m == n == o):
            result.append(m)
    return result
\end{lstlisting}
The constrained solution:
\begin{lstlisting}[style=python, frame=none]
def extract_index_list(l1, l2, l3):
    result = []
    index = 0
    min_len = min(len(l1), len(l2), len(l3))
    while True:
        if index >= min_len:
            break
        m, n, o = l1[index], l2[index], l3[index]
        if m == n == o:
            result.append(m)
        index += 1
    return result
\end{lstlisting}

    \item Positive value: The instruction led to a solution that is educationally valuable, algorithmically alternative, or engineering-robust. Although the new solution may not necessarily be better than the original one, it demonstrates a different way of thinking. The vanilla solution: 
\begin{lstlisting}[style=python, frame=none]
def permutation_coefficient(n, k): 
    P = [[0 for i in range(k + 1)] for j in range(n + 1)] 
    for i in range(n + 1): 
        for j in range(min(i, k) + 1): 
            if (j == 0): 
                P[i][j] = 1
            else: 
                P[i][j] = P[i - 1][j] + (j * P[i - 1][j - 1]) 
            if (j < k): 
                P[i][j + 1] = 0
    return P[n][k] 
\end{lstlisting}
The constrained solution:
\begin{lstlisting}[style=python, frame=none]
from functools import reduce
def permutation_coefficient(N, K):
    if K == 0:
        return 1
    if K > N:
        return 0
    return reduce(lambda X, Y: X * Y, range(N, N - K, -1))
\end{lstlisting}

\end{itemize}
\end{itemize}

\section{Prompt}\label{app:prompt}

\begin{PromptBlueBox}{Actor prompt}
As a programming assistant, your task is to generate code snippets based on the user question and instructions given below:
\newline

\#\# Requirements

- Make sure you follow the user instructions. If the instruction says to use a specific language or a specific method, use exactly that. 
\newline
- Your output should be a valid code snippet in the programming language indicated in the question or the instructions.
\newline
- Remember to import any necessary libraries or modules if needed.
\newline
- Remember to import Typing if the signature contains type hints.
\newline

\#\# Output Format

The output should only be a valid code snippet without any explanations, comments, or text outside the code.
\newline

\#\# Problem

\{prompt\}
\end{PromptBlueBox}

\begin{PromptBlueBox}{LLM evaluator for instruction following}
\# Role
\newline
You are an expert Code Compliance Auditor. Your task is to verify whether a model-generated code solution strictly follows a specific set of user instructions.
\newline
\newline
\# Workflow
\newline
You must follow these two steps strictly:
\newline
\newline
\#\# Step 1: Step-by-Step Analysis
\newline
- Iterate through each instruction in the provided list.
\newline
- For each instruction, examine the "Model-generated Code" to see if the requirement is met.
\newline
- Briefly explain your reasoning for each instruction.
\newline
- Conclude each analysis with a "Yes" or "No".
\newline
\newline
\#\# Step 2: Final Output Generation
- Collect your "Yes" or "No" verdicts into a Python-style list.
\newline
- Ensure the list length is exactly the same as the number of instructions.
\newline
- Wrap this list inside `<answer>' and `</answer>' tags.
\newline
- The content inside `<answer>' must be **only** the list (e.g., `[`Yes', `No']'), without any reasoning text.
\newline
\newline
\# Evaluation Criteria
\newline
- **Strict Adherence:** If a specific language, library, algorithm complexity, or naming convention is requested, it must be present.
\newline
- **Syntax:** The code must be syntactically valid for the requested language.
\newline
- **Order:** The $i$-th element in your result list must correspond to the $i$-th instruction.
\newline
\newline
\# Input Data
\newline
\#\# Coding Question
\newline
\{question\}
\newline
\newline
\#\# Instructions
\newline
\{instruction\}
\newline
\newline
\#\# Model-generated Code
\newline
\{code\}
\newline
\newline
\# Execution
\newline
Please begin your Step 1 Analysis now, followed immediately by the Step 2 Final Output.
\end{PromptBlueBox}

\begin{PromptBlueBox}{Proof-by-construction by generator}
You are an expert Python coding challenge designer and data synthesizer. Your task is to "mutate" a given programming problem (Seed Question) by applying a specific **instruction**, thereby generating a new, more challenging version of the problem.
\newline
\newline
\# Input Format
\newline
You will receive an XML snippet containing:
\newline
1. `<question>': The original problem description, existing instructions, the reference solution (Code), programming language, and language.
\newline
2. `<Mutations>': A list of `<Mutation>' tags. Each mutation is a template for a new instruction and a list of `<params>' required to instantiate that template.
\newline
\newline
\# Workflow (Step-by-Step)
\newline
Before generating the final XML output, you must perform the following reasoning steps inside a `<thought>' tag:
\newline
1. Analyze the Original Code:
\newline
- Understand the algorithm, complexity, and existing instructions of the seed code.
\newline
\newline
2. Strategic Parameter Selection and Applicability Check:
\newline
- Iterate through **each** provided `<Mutation>' in the list.
\newline
\text{\ \ \ \ }- For each candidate, evaluate two criteria:
\newline
\text{\ \ \ \ }\text{\ \ \ \ }- **Compatibility**: Does this mutation make sense for this problem and the current instructions?
\newline
\text{\ \ \ \ }\text{\ \ \ \ }- **Challenge Level**: How much does this force a refactor?
\newline
\text{\ \ \ \ }- **Selection Strategy**:
   \newline
\text{\ \ \ \ }\text{\ \ \ \ }- Discard Incompatible mutations.
     \newline
\text{\ \ \ \ }\text{\ \ \ \ }- If multiple have the same challenge level, prioritize the one that aligns best with Pythonic best practices or specific algorithmic concepts.
\newline
\newline
3. Parameter Instantiation:
\newline
- For the selected mutation, look at its `<params>'.
\newline
\text{\ \ \ \ }- If `<Mutation>/<params>' is empty, skip parameter selection and proceed directly to conflict detection. The instruction text is fixed.
\newline
\text{\ \ \ \ }- If `<Mutation>/<params>' is not empty:
\newline
\text{\ \ \ \ }\text{\ \ \ \ }- **Maximize Challenge**: Choose parameter options that contradict the *current* implementation (e.g., if code uses a `for` loop, and options are `['while', 'recursion']`, choose `recursion` if valid).
\newline
\text{\ \ \ \ }\text{\ \ \ \ }- **Feasibility**: Ensure the chosen parameters allow for a valid solution.
\newline
\newline
4. Conflict Detection:
\newline
- Ensure the selected mutation and its parameters do not contradict the original `<instruction>'.
\newline
- Such as 'Your code must utilize exactly 1 list comprehension.' and 'Your code must not use any for loops.', which are incompatible with each other.
\newline
\newline
5. Refactor Code (Only if Successful):
\newline
- If successful, rewrite the reference code to **strictly adhere** to the new instantiated constraint.
\newline
- The modified code must produce the exact same output for the same inputs as the original code.
\newline
\newline
6. Synthesize Output:
\newline
- Generate the `<success>' tag first.
\newline
\text{\ \ \ \ }- If `<success>false</success>', **STOP** after closing the tag. Do not generate params or question.
\newline
\text{\ \ \ \ }- If `<success>true</success>', generate `<instantiated\_params>' and the modified `<question>'.
\newline
\text{\ \ \ \ }\text{\ \ \ \ }- If the input `<Mutation>/<params>' was empty, the tag <instantiated\_params> must also be empty (i.e., `<instantiated\_params></instantiated\_params>').
\newline
\newline
\# Output Format

Return the result strictly in XML format. Do not include markdown code block markers (like ```xml).

\begin{lstlisting}[language=xml]
<output>
  <thought><![CDATA[1. Analysis: [Analysis of original code]
2. Evaluation:
   - Mutation ID 1: [Compatibility: Yes/No] | [Challenge: Low/Med/High] | [Reasoning]
   - Mutation ID 2: ...
   - Decision: Selected Mutation ID [X] because...
3. Parameter Selection: [Reasoning for chosen params]
4. Refactoring Strategy: [How the code will change]]]></thought>
  <success>[true/false]</success>
  <instantiated_params>
    <param>
      <name>[Parameter Name, e.g., variable_name]</name>
      <value>[Selected Value, e.g., total_score]</value>
    </param>
  </instantiated_params>
  <question>
    <question_desc>[Original description. Modify only if the mutation fundamentally changes the output format, otherwise keep as is.]</question_desc>
    <instruction>
  [Existing instructions]
  [New instantiated instruction]
    </instruction>
    <code><![CDATA[
  [The full, modified Python code]
    ]]></code>
    <function_signature><![CDATA[ function signature that should be aligned with the code ]]></function_signature>
    <lg>[en/zh]</lg>
    <programming_language>python</programming_language>
  </question>
</output>
\end{lstlisting}

\# Constraints
\newline
- Do **not** alter the core algorithmic intent (e.g., if the problem is "Sum of list", do not change it to "Average of list").
\newline
- The `<instruction>' field must contain **all** previous instructions plus the new one, separated by newlines and numbering.
\newline
- The code must be wrapped in `<![CDATA[ ... ]]>'.
\end{PromptBlueBox}

\begin{PromptBlueBox}{Instruction mutation}
You are an expert in program analysis and synthetic coding data generation.
\newline
\newline
Your task is to EVOLVE an existing instruction into a STRICTLY STRONGER, AST-CHECKABLE instruction that INVALIDATES ALL GIVEN EXAMPLES, while still SUBSUMING the original instruction.
\newline
\newline
The goal is to generate a higher-difficulty, verifier-backed instruction that exposes genuine unconstrained syntactic degrees of freedom.
\newline
\newline
\#\# Input
\newline
You will receive an XML describing an existing instruction, including:
\newline
  - type
\newline
  - instruction template
\newline
  - checking function
\newline
\newline
Multiple XML examples of solved coding problems, containing:
\newline
  - question\_desc
\newline
  - instruction(s)
\newline
  - example\_input\_and\_output
\newline
  - code
\newline
  - instantiated\_params
\newline
\newline
All provided examples are VALID solutions to the ORIGINAL instruction.
\newline
\newline
\#\# Goals
\newline
You MUST design an EVOLVED instruction such that:
\newline
\newline
1. Subsumption
\newline
   - Any program that satisfies the EVOLVED instruction MUST also satisfy the ORIGINAL instruction.
\newline
   - You must explicitly justify this subsumption.
\newline
   - The evolved instruction MUST introduce at most ONE new AST-level constraint beyond those already enforced by the original instruction.
\newline
    - The evolved instruction shouldn't be overly complicated
\newline
\newline
2. Examples-as-Negative
\newline
   - ALL provided example programs MUST FAIL the EVOLVED instruction.
\newline
   - These failures must arise from a GENERAL, STRUCTURAL, AST-level constraint.
\newline
   - You MUST NOT reference specific identifiers, literals, or fingerprints unique to the examples.
\newline
\newline
3. AST-Checkability
\newline
   - The evolved instruction MUST be checkable using static AST analysis.
\newline
   - No runtime execution, no I/O, no performance, no semantic reasoning.
\newline
\newline
4. Generalization
\newline
   - The evolved instruction MUST generalize beyond the given examples.
\newline
   - You are FORBIDDEN from writing constraints that merely exclude the examples without structural meaning.
\newline
\newline
You MUST NOT:
\newline
- Mention or encode specific variable names, constants, or literals that appear only in the examples.
\newline
- Refer to line counts, whitespace, formatting, or comments.
\newline
- Use vague or semantic constraints such as: "readable", "clean", "efficient", "reasonable", "well-structured".
\newline
- Encode example-specific signatures or hashes.
\newline
- Introduce constraints that can be satisfied or violated trivially.
\newline
\newline
\#\# Output format (MUST be valid XML, no extra text):
\newline
The output should only be a valid XML document without any extra text:
\begin{lstlisting}[language=xml]
```xml
<output>
  <thought><![CDATA[
    A detailed explanation covering:
    - Interpretation of the original instruction
    - How the examples satisfy it
    - What syntactic freedom is being restricted
    - Why the evolved instruction is strictly stronger
    - Why it subsumes the original
    - Why all provided examples fail it
    ]]></thought>

  <instruction><![CDATA[
    The evolved instruction template (may contain placeholders)
  ]]></instruction>

  <function>
    <name>FUNCTION_NAME</name>
    <params>
      <name>param_name</name>
      <type>string | integer | boolean | enum</type>
      <options>...</options> <!-- only if type=enum -->
    </params>
    <params>
      <name>param_name</name>
      <type>string | integer | boolean | enum</type>
      <options>...</options> <!-- only if type=enum -->
    </params>
    ...
    <impl><![CDATA[
def FUNCTION_NAME(tree, code_str, param1, param2, **kwargs):
    # AST-based deterministic checking logic
    return True or False
]]>
    </impl>
  </function>

  <positive_cases>
    <case>
      <instantiated_params>
        <!-- a valid parameter instantiation -->
      </instantiated_params>
      <code>
      <![CDATA[
# Python code that SATISFIES the evolved instruction under the given parameter instantiation
]]>
      </code>
    </case>
    <case>
      <instantiated_params>
        <!-- a valid parameter instantiation -->
      </instantiated_params>
      <code>
      <![CDATA[
# Python code that SATISFIES the evolved instruction under the given parameter instantiation
]]>
      </code>
    </case>
  </positive_cases>
</output>
```
\end{lstlisting}
\end{PromptBlueBox}

\begin{PromptBlueBox}{Problem rephrase}
You are an expert Data Augmentation Specialist for code generation tasks. Your goal is to rewrite coding instructions to increase linguistic diversity while STRICTLY preserving the original semantic meaning and logic.
\newline
\newline
\# Task
\newline
You will receive a list of coding instructions (formatted as a numbered list). You must generate a rephrased version of this list. The new instructions should sound natural, diverse, and human-written.
\newline
\newline
\# Transformation Rules
\newline
1. **Rephrase**: This is your main tool. Use synonyms, change sentence structures, and vary the tone to express the exact same requirement.
\newline
2. **Combine**: Merge two or multiple instructions into a single instruction.
\newline
3. **Preserve Logic**: 
\newline
   - Do NOT change any values (e.g., if it says "length < 5", do not output "length <= 5").
\newline
   - Do NOT change the requirement type (e.g., "must use" cannot become "can use").
\newline
   - Do NOT omit any constraints.
\newline
4. **Format**: The output must be a numbered list separated by newlines.
\end{PromptBlueBox}

\end{document}